\documentclass[12pt]{article}

\usepackage{amsthm}
\usepackage{mathtools}
\DeclareMathAlphabet{\mathbbold}{U}{bbold}{m}{n}

\usepackage{tikz}
\usetikzlibrary{shapes.geometric, positioning, arrows, animations, calc}
\tikzstyle{process} = [rectangle, draw, minimum width=3.5cm, minimum height=1cm]
\tikzstyle{processn} = [rectangle, draw=none, minimum width=3.5cm, minimum height=1cm]
\tikzstyle{decision} = [diamond, draw, minimum width=3cm, minimum height=1cm]
\tikzstyle{arrow} = [thick,->,>=stealth]
\usepackage{makecell}

\usepackage{adjustbox} 

\usepackage[T1]{fontenc}
\usepackage[utf8]{inputenc}
\usepackage{newtxtext, newtxmath}
\usepackage[protrusion=true, expansion=true]{microtype}
\usepackage{setspace}

\usepackage[english]{babel}
\usepackage{csquotes}

\usepackage[margin=1in]{geometry}

\usepackage{booktabs}
\usepackage{multirow}
\usepackage[hang, small, labelfont=bf, textfont=it, up]{caption}
\usepackage{enumerate}
\usepackage[capposition=top]{floatrow}
\usepackage{graphicx}
\usepackage{longtable}
\usepackage{threeparttable}
\usepackage{threeparttablex} 

\usepackage{abstract}

\usepackage{titlesec}
\titleformat*{\section}{\large\bfseries}
\titleformat*{\subsection}{\normalsize\bfseries}

\usepackage[hidelinks]{hyperref}

\usepackage{pdflscape}

\definecolor{darkred}{RGB}{100, 10, 44}
\definecolor{darkgreen}{RGB}{25, 60, 20}
\definecolor{darkling}{RGB}{58, 66, 66} % Deep Brown
\definecolor{darktheme}{RGB}{61, 97, 97}
\definecolor{bgreen}{RGB}{80, 150, 100}
\definecolor{orange}{HTML}{FFAA33} 
\definecolor{bblue}{RGB}{102, 178, 178}

\usepackage[authordate, backend=biber]{biblatex-chicago}
\usepackage[most]{tcolorbox} 

\usepackage{titling}

\title{
{\Large\bfseries{
The Effects of Flipped Classrooms in Higher Education:\\ A Causal Machine Learning Analysis}\thanks{Funded by the German Federal Ministry of Education and Research. Grant reference: 16DHL1037. The grant provider was not involved in the design, creation, analysis, evaluation, or submission of this work in any way. We thank Sonja Schattevoy and Carolin Walz for excellent research assistance. 
We gratefully acknowledge the Multimedia Center of Heinrich Heine University Düsseldorf, and especially Tobias Koch, for implementing the tracking functionality to collect detailed data on video-watching  behavior.
}
}
}
\author{
\normalsize
Daniel Czarnowske
\thanks{
Heinrich Heine University Düsseldorf, Chair of Statistics and Econometrics, Universitätsstraße 1, 40225 Düsseldorf, Germany; e-mail: \texttt{\href{mailto:daniel.czarnowske@hhu.de}{daniel.czarnowske@hhu.de}}
}
\and
\normalsize
Florian Heiss
\thanks{
Heinrich Heine University Düsseldorf, Chair of Statistics and Econometrics, Universitätsstr. 1, 40225 Düsseldorf, Germany; e-mail: \texttt{\href{mailto:florian.heiss@hhu.de}{florian.heiss@hhu.de}}
}
\and
\normalsize
Theresa M. A. Schmitz
\thanks{Heinrich Heine University Düsseldorf, Chair of Statistics and Econometrics, Universitätsstr. 1, 40225 Düsseldorf, Germany; e-mail: \texttt{\href{mailto:theresa.schmitz@hhu.de}{theresa.schmitz@hhu.de}}}
\and
\normalsize
Amrei Stammann
\thanks{University of Bayreuth, Chair of Empirical Economics, Universitätsstraße 30, 95447 Bayreuth, Germany; e-mail: \texttt{\href{mailto:amrei.stammann@uni-bayreuth.de}{amrei.stammann@uni-bayreuth.de}}}
}

\date{\small\today}

\begin{document}

\maketitle

\thispagestyle{empty}

\renewcommand{\abstractname}{\vspace{-5em}}
\begin{abstract}
\noindent

This study uses double/debiased machine learning (DML) to evaluate the impact of transitioning from lecture-based blended teaching to a flipped classroom concept. Our findings indicate effects on students' self-conception, procrastination, and enjoyment. We do not find significant positive effects on exam scores, passing rates, or knowledge retention. This can be explained by the insufficient use of the instructional approach that we can identify with uniquely detailed usage data and highlights the need for additional teaching strategies. Methodologically, we propose a powerful DML approach that acknowledges the latent structure inherent in Likert scale variables and, hence, aligns with psychometric principles.

\footnotesize
\vspace{1em}
\noindent\textbf{Keyword:} Causal Machine Learning, Learning Analytics, Flipped Classroom in Higher Education, Introductory Statistics Course\\
\noindent\textbf{JEL:} A20, A22, I20, I21, I23
\end{abstract}

\onehalfspacing

\newpage

\section{Introduction} \label{s1_intro}

Advances in technology are transforming the landscape of higher education, offering instructors a wide range of tools and teaching formats to improve their instruction \parencite{escueta2020}.
The flipped classroom is a prevalent student-centered instructional format that inverts traditional teaching methods and combines them with appropriate technologies, such as videos, online quizzes, and live voting. In particular, students engage with new course content before class, while traditional home activities are carried out in the classroom. As a result, the instructor becomes a facilitator, enabling students to study the content at their own pace.\footnote{For more details, see, among others, \textcites{strayer, o_flaherty, fleenor}.} By structuring learning around spaced, active self-learning before class and in-class knowledge deepening activities, flipped classrooms aim to improve student learning outcomes and long-term retention \parencite{mc_kenzie} and therefore human capital accumulation. The flipped curriculum involves sequential phases of independent knowledge acquisition through (typically digital) materials, followed by classroom sessions designed for active application and resolution of misconceptions \parencites{davies, kim}. Given the growing and more diverse student populations in higher education, characterized by different backgrounds and prior knowledge, the flipped classroom approach theoretically holds promise to better address individual learning needs than traditional lectures \parencites{maringe, gannaway}.

Despite this theoretical advantage, empirical studies in higher education do not reveal strong evidence that the flipped classroom format is superior to traditional lecture-based settings (for recent overviews, see \cites{akccayir, cheng, o_flaherty}). This could be due to reliance on self-reported or inadequate data \parencites{o_flaherty, flipps_mainz}. Furthermore, several studies lack a solid theoretical foundation or do not completely align with the flipped classroom model \parencites{o_flaherty}. Some compare instructional environments with significant structural changes or allow self-selection \parencites{bintz, strayer, cheng}, two problems that are prevalent in quasi-experimental designs like cohort studies. Moreover, essential details of the design and environment are often not reported in sufficient detail \parencite{cheng}. 

This study uses a cohort design to assess the effect of changing the instructional format of a large, compulsory introductory statistics course for business and economics students at a tuition-free German university from lecture-based (instructor-centered) teaching to a flipped classroom. To address the potential confounding inherent in this observational study, we collected data on learning-theoretically relevant covariates (sociodemographics, learning preconditions, perception of statistics, and learning strategies) using paper questionnaires. Given that the number of additional covariates collected is quite large (relative to our sample size), and the fact that these covariates are highly correlated, we employ the Double Machine Learning (DML) approach of \textcite{chernozhukov} in combination with suitable machine learning (ML) techniques to have valid inference in our high-dimensional setting. Although students indicated improvements in their self-concept and a decrease in procrastination as a result of changing the instructional format, their enjoyment of the course decreased. Contrary to the expectation that flipped classrooms produce positive learning outcomes, we did not find significant positive effects in our study. However, detailed usage data reveal that the average student is not implementing the flipped classroom curriculum in a sufficient way, highlighting the need to provide complementary strategies to ensure that students can benefit from the instructional format. This is in line with several other studies, such as, for example, \textcite{al-zahrani2015}, which report inadequate student preparation as one of the main difficulties associated with flipped classrooms (also see \textcite{akccayir} for a systematic review of the literature on the advantages and challenges of the flipped classroom).

This study makes the following contributions. First, within the flipped classroom literature, our careful implementation of a flipped classroom and comprehensive data collection allows us to obtain reliable estimates of the change in the instructional format on various student outcomes. In addition, leveraging granular usage data from the flipped curriculum, we provide nuanced insights into students' patterns of engagement with the learning materials. Second, we demonstrate a powerful and objective methodological approach by successfully employing DML combined with ML techniques to (implicitly) reduce the dimensionality of a large set of covariates (including answers to 67 Likert-type questions) for application in studies with similar research designs. Furthermore, we discuss suitable ML techniques, such as ridge regression, that align with psychometric principles, acknowledging the latent factor structure inherent in Likert scale variables, such as tendencies to procrastinate. 

Several empirical studies have examined the impact of transitioning to a flipped classroom format on student achievement in economics programs or courses. Some of these studies employ cohort comparisons to assess the effect of the flipped classroom on student performance (see, for example, \textcites{lage2000, findlay2014, swoboda2016}). \textcite{swoboda2016}, for example, analyze the effect of a flipped classroom in an introductory microeconomics course using a cohort design with a total of 150 students. They find a significant positive effect of the flipped format on the difference between students' pre- and post-test scores on a comprehension test. However, unlike our study, they use scores from a voluntary, unsupervised test instead of actual exam results as outcome variable. Additionally, their analysis includes a more limited set of controls for socio-demographic characteristics and prior learning conditions, and does not incorporate any psychometric variables. This is a general limitation of many observational studies. Few studies employ randomized controlled trials (RCT) to investigate the effects of the flipped classroom on student outcomes (see, for example, \textcites{wozny2018a,setren2021}). Although RCTs are known for their robust causal identification, their implementation often entails substantial costs \parencites{wozny2018a, wozny2018b}. Furthermore, ethical and legal constraints frequently preclude the execution of RCTs. This is true in our study where student assignment to specific course formats through randomization was not permissible. Despite these limitations, we are convinced that our rich set of control variables combined with suitable machine learning techniques adequately addresses potential selection concerns, thereby yielding reliable results within our quasi-experimental design.

In addition to research on flipped classrooms, there are several studies that examine the impact of various technologies often integrated into flipped classroom models on student outcomes (see, for example, \textcites{figlio2013,bowen2014,joyce2015,alpert2016,bettinger2017,cacault2021,depaola2023, kofoed2024}).

The subsequent sections of this paper are structured as follows. Section \ref{s2_ins} details the cohort design and describes the course structure before and after the change in instructional format. Section \ref{s3_data} describes and provides summary statistics for the data. Section \ref{s4_meth} presents our estimation approach. Section \ref{s5_res} reports the estimated effects of the change in the instructional format on various student outcomes. Section \ref{s6_discussion} offers a discussion of the mechanism driving our null results for exam performance and knowledge retention. Section \ref{s7_concl} concludes.

\section{Cohort Design} \label{s2_ins}

We compare two subsequent cohorts from a large introductory statistics course at a public university in western Germany, each attracting more than 500 students. This is a compulsory course typically taken in the first semester of the Bachelor's degree programs in Economics and Business Administration, though students may choose to take it later in their curriculum. The course is taught over one semester and covers the fundamentals of descriptive statistics. In addition, the course introduces students to the statistical software R \parencite{r_software} and the graphical user interface RStudio \parencite{rstudio_software}. Students learn both theoretical methods and how to apply them using R. This course is well suited to assess the impact of a change in instructional format because its large enrollment improves data quality. In addition, the benefits of the inverted classroom approach are potentially high given the heterogeneous student population in terms of prior knowledge, interest, and talent in math and statistics, as well as computer literacy. Therefore, the course offers a particularly suitable setting for studying the effects of the flipped classroom format on students' learning outcomes and perceptions. The treatment consists of an entire change in the teaching concept (involving several components), and the change was not announced to the students to avoid self-selection from the first to the second cohort. The first cohort (Cohort 1) was taught in the winter term 2017/2018 in a lecture-based teaching setting, while the course design for the second cohort (Cohort 2) during the winter term 2018/2019 followed a flipped classroom approach. In the following, we describe the similarities and differences of both instructional formats in more detail.

The instruction in both cohorts was based on the same lecture slides, but the knowledge acquisition channel differed. Students in the lecture-based cohort first came into contact with the content during classical 90-minute lectures at the university, while students in the flipped classroom cohort were expected to acquire the content by themselves watching specifically tailored instructional videos given a provided weekly schedule. These videos were based on the sub-chapters of the lecture slides, on average 14 minutes long, summing to roughly 70 minutes of video content per week. The instructional videos were designed to include active learning phases in which students are supposed to pause viewing to solve problems. This design aimed to maintain an overall roughly equivalent workload of 90 minutes for knowledge acquisition in both the flipped and the lecture-based formats. The videos remained accessible throughout the semester once they were uploaded to the university's streaming platform. To encourage students to engage with the instructional videos in time, we offered voluntary weekly online quizzes based on the self-taught content of the corresponding videos that allowed them to collect up to six bonus points for the final exam. Each online quiz had a one-day processing time, and after the deadline, the students received automatic feedback on their results. In addition, after each quiz day, students had instructor-guided classroom sessions with clicker polls and discussions scheduled to gain a deeper understanding of the content and clarify remaining questions. To benefit from these classroom sessions and really deepen understanding, it is crucial that students attend well prepared. Active participation in the classroom session required that the students have engaged with the scheduled video content in time and ideally already tested that self-taught knowledge and identified minor misconceptions. 

For the lecture-based cohort, we also offered three (voluntary) online quizzes, each with a processing time of several days, in which students could earn up to three bonus points if they successfully completed all quizzes. Unlike the flipped classroom cohort, these online quizzes only covered questions on the content that the lecturer and the tutors presented so far. Furthermore, recordings of the lecture were made available ex post. Importantly, these 90-minute recordings were not specifically prepared instructional videos as those for the flipped cohort. In both cohorts, students further deepened their knowledge in the weekly tutorials offered. For both cohorts, these tutorials took place in person on campus, were not recorded, and used the same material. At the end of the semester, students in both cohorts wrote a cumulative final exam with a maximum score of 60 points. For better comparability, we did not change the exam questions, but we ensured that the exam from the first cohort was not accessible to students from the second cohort. Furthermore, the same exams or similar questions were  never used in any exams before, ruling out anticipation effects. In the following semester, an advanced compulsory statistics course is part of the curriculum. For both cohorts, we conducted an unannounced knowledge retention test consisting of selected exam tasks from the introductory course in its first lecture. Since it was not possible to administer the test under typical exam conditions, where students are spread over multiple rooms to maintain spacing and prevent cheating, we used two different versions of the test.

In summary, the recommended workflow for the first cohort involved attending the lecture first, followed by the tutorial, and then completing the corresponding online quiz. In the second cohort, the recommended workflow required students to engage during self-learning phases first with the same lecture slide content through videos and then taking the online quizzes, before participating in active classroom sessions and finally attending the tutorial.

\section{Data} \label{s3_data}
 
We collect detailed individual-level data from several paper questionnaires and performance data from the final exam and the knowledge retention test.\footnote{{The paper questionnaires were designed in coordination with colleagues from the University of Mainz, see, among others, \textcite{flipps_mainz}.}} The questionnaires were answered in the first and last lecture/active classroom session. Participation in the questionnaires was voluntary, but participants were encouraged by the opportunity to win an Apple Ipad. From the first questionnaire, we obtained self-reported information on sociodemographic characteristics, such as age, sex, or migration background, as well as learning preconditions, such as the study program, current semester of study, or grade in math in high school diploma. In addition, we used Likert scale questions to assess students’ learning strategies and their perception of statistics in general. In the second questionnaire, other Likert scale questions were included to learn about students' perceptions of lectures/active classroom sessions. These questions were developed based on established instruments and scales from the psychometric literature (see \textcite{schau}, \textcite{pintrich}, \textcite{pintrichQ}, \textcite{pekrun}, \textcite{patzelt}). An overview of the latent constructs measured and their corresponding scales is provided in the Online Appendix \ref{app:likert_scales}. For both cohorts, we also know the total points that students achieved in the final exam (without bonus points) and in the knowledge retention test.

We restrict our analysis to students in their first semester, who were registered for the final exam and for whom we have complete data. Our main data set consists of 420 students, 202 in the lecture-based cohort, and 218 in the flipped classroom cohort (Sample A). We are using this entire sample to study the effects of changing the instructional format on exam points and pass rates. For our analyses on the effect on knowledge retention and on students' perception and learning strategies, we formed two subsets of this main data set, since not every student who took the final exam participated in the knowledge retention test and the second questionnaire. We therefore constructed a subsample of 310 individuals (144 in Cohort 1, 166 in Cohort 2) to analyze knowledge retention (Sample B) and a subsample of 226 individuals (96 in Cohort 1, 130 in Cohort 2) to analyze students' perception and learning strategies (Sample C). 

Table \ref{tab:sociodemographics} presents the descriptive statistics for the sociodemographic variables. In all three samples and both cohorts, respectively, the majority of students are between 17 and 19 years old, do not look after children or care for relatives, graduated in the German federal state North-Rhine Westphalia (NRW), are native German speakers without a migration background, and do not work alongside studying. Female and male students are roughly equally represented in all samples and cohorts.

Table \ref{tab:learningpreconditions} summarizes the descriptive statistics for the variables of the learning preconditions. More than 95\% of the students achieved the university entrance qualification with the \textit{Abitur} (German high-school diploma), and most of them obtained it from the \textit{Gymnasium} (German high-school). The average final school grade ranges from 2.675 to 2.908, where 1 is the lowest passing grade and 4 is the best possible grade.\footnote{The usual grade system in Germany is in reversed order, i.e., 1 (very good), 2 (good), 3 (satisfactory), 4 (pass), 5 (poor), 6 (insufficient). One can only get \textit{Abitur} with a grade equal to or better than 4.} Furthermore, the majority of students are enrolled in business administration, did not take an advanced mathematics course in high school, and achieved a good or satisfactory grade in their most recent high school mathematics course. Almost all of the students have no previous study experience. Note that even though we restricted our sample to first semester students, some students studied a different subject before enrolling into business administration or economics. This also partly explains why a small fraction of students already attended other university statistics and mathematics courses. Another possible explanation is that some students participated in preparatory courses shortly before starting their studies or attended university discovery days. Most students have not completed a vocational training prior to studying. Some students have already gained experience working with statistical software, which may have been part of a high school course, vocational training, or during prior enrollment in another study program. We also asked the students what grade they were aiming for in the final exam and what grade they think they would achieve. Interestingly, we observe a discrepancy between students' expectations and aspirations. Although most students either expected a good or satisfactory outcome, they aimed for a very good and good outcome, reflecting ambition, but also realism.

In the first lecture or active classroom session, we asked students with several 7-point Likert scale questions about their ex ante perception of the subject statistics and their learning strategies. In table \ref{tab:likertq1}, we report the means of all Likert items that form a certain latent construct. For example, the latent construct \textit{affection} consists of six questions on the Likert scale. A value of -3 indicates a distinctly negative attitude toward statistics, while the maximum value of 3 reflects a clearly positive attitude. A value of zero represents a neutral stance towards the subject. For all latent constructs that describe the perception of statistics, we only find positive means throughout all cohorts and samples. Therefore, on average, students in our analyses rather showed openness towards statistics (\textit{affection}), consider statistics as difficult (\textit{difficulty}), plan to work hard for the course (\textit{effort}), have ambitions to achieve a good grade in the course (\textit{extrinsic motivation}), are interested in the subject (\textit{interest}), are optimistic that they will master the course (\textit{self-concept}), and believe that statistics is useful for their further studies, job, or daily life (\textit{value}). For the following latent constructs, representing students' learning strategies, we find positive averages. On average, students tend to be able to critically reflect on the course material (\textit{critical thinking}), appear to be rather well organized in planning their learning phases (\textit{environment}), use repetition to learn (\textit{repetition}), engage in collaborative learning with peers (\textit{peer}), and are more capable of monitoring and adapting their learning progress (\textit{regulation}). Furthermore, we find negative means for the latent constructs \textit{procrastination} and \textit{text-anxiety} meaning that our average students tend to procrastinate less when learning and are not particularly anxious about the exam.

To ensure comparability, we assess whether both cohorts have similar distributions of observed covariates. In particular, we use mean comparison tests. The tests are reported in Tables \ref{tab:sociodemographics}, \ref{tab:learningpreconditions}, and \ref{tab:likertq1} and confirm that students in both cohorts are on average quite similar (at a significance level of 5\%). Regarding sociodemographic characteristics, we only find statistically significant differences in the age distribution in Sample B. For learning preconditions, the type of high-school significantly differs in Sample A. In Sample B, last high school math grade and the enrollment in vocational training show significant differences, while the reversed high school grade is significantly different in Sample C. Additionally, for Sample A and B, the latent construct \textit{value} exhibits a significantly higher average in the flipped classroom than in the lecture-based cohort.

Table \ref{tab:outcomes} summarizes the descriptive statistics of the variables that we use as our outcome variables. We present only the statistics for the sample that was used to estimate the effect of the change in the instructional format on the respective outcome variable. The average student in Cohort 1 achieved 28.146 points in the final exam and in Cohort 2 roughly 27.456. Although we used the same threshold for passing the exam in both cohorts, we find a slightly higher passing rate of 80.2\% in Cohort 1 compared to a passing rate of 74.3\% in Cohort 2. Neither the differences in exam points nor in passing rates are statistically significant. The students also performed similarly on the knowledge retention test, as in both cohorts they achieved roughly 5.5 points out of 8. For the latent outcome variables, we find the following four statistically significant differences between both cohorts. Recall that all of these variables were collected with a paper questionnaire in the last lecture/active classroom session. On average, students in the lecture-based cohort were less optimistic about their ability to master statistics at the end of the course than students in the flipped classroom cohort. Moreover, at the end of the course, on average, students in Cohort 2 procrastinated learning less than students in Cohort 1. We also found that after attending the entire course, the average student in the flipped classroom format was more optimistic about mastering the subject in classroom sessions than the average student in the lecture-based format. An unfortunate finding is that, on average, students in Cohort 2 enjoyed the active classroom sessions less than students from Cohort 1 enjoyed the lecture.

	\begin{landscape}
		\begin{table}[!htbp]
            \begin{threeparttable}
			\centering
			\footnotesize
			\caption{Sociodemographics}
            \label{tab:sociodemographics}
			\begin{tabular}{lccccccccc}
				\toprule
				& \multicolumn{3}{c}{Sample A} 
				& \multicolumn{3}{c}{Sample B} 
				& \multicolumn{3}{c}{Sample C} \\
				\cmidrule(lr){2-4} \cmidrule(lr){5-7} \cmidrule(lr){8-10}
				Variable
				& Cohort 1 & Cohort 2 & p-value
				& Cohort 1 & Cohort 2 & p-value
				& Cohort 1 & Cohort 2 & p-value \\
				\midrule
				\underline{Age} &  &  & 0.297   &  & & 0.042  &  & & 0.114 \\
				\hspace{.1cm} $20-22$  &   0.292 & 0.307  &   & 0.264  &  0.301 &  &   0.208   &  0.300 &  \\
				\hspace{.1cm} $23 +$  & 0.124  &  0.078  &   &   0.146 & 0.060  &  & 0.125 & 0.062&  \\
				%		\midrule
				Caregiver   &   0.025 & 0.041  &   0.347 &   0.021 & 0.042 &   0.290 &  0.021  & 0.038 &  0.452 \\
				%		\midrule
				\underline{Federal State}  &  &  &   0.453  &  & & 0.843  &  & & 0.932  \\
				\hspace{.1cm} Foreign Country  &  0.030 & 0.014   &   & 0.028           &  0.018 &  &     0.031    &    0.023 &  \\
				\hspace{.1cm} NRW   &   0.891 &  0.890     &   & 0.875 & 0.880  &  &  0.885 & 0.892 &  \\
				%		\midrule
				Female &  0.455 &  0.459     & 0.947   &   0.507 &             0.482  & 0.662 &    0.542  & 0.500 & 0.538 \\
				%		\midrule
				\underline{Migration Parents}  &  &  & 0.560   &  & & 0.621  &  & &   0.189 \\
				\hspace{.1cm} Both German   &  0.599 & 0.628   &   & 0.653 &             0.633 &  & 0.635 & 0.685 &  \\
				\hspace{.1cm} Father German   & 0.114 & 0.073   &   & 0.104 &             0.072 &  &  0.135    &  0.054 &  \\
				\hspace{.1cm} Mother German   &  0.050  & 0.055  &   & 0.035 & 0.048 &  &   0.052 & 0.046 &  \\
				%\midrule
				Mother tongue German &   0.822 &  0.803 & 0.619   &   0.854 &           0.801 & 0.221   &  0.875 & 0.862 &  0.769 \\
				%		\midrule
				\underline{Work (in hours per week)}  &  &  & 0.943     &  & &  0.924  &  & & 0.771 \\
		    	\hspace{.1cm} $(0 - 5]$  &  0.064  &  0.078  &   &   0.056  & 0.072 &  &    0.052  &  0.069 &  \\
				\hspace{.1cm} $(5 - 10]$ & 0.252 &  0.239  &   &  0.222  &           0.205  &  &  0.250  &  0.223 &  \\
				\hspace{.1cm} $>10$  &   0.144 &  0.138  &   &   0.146 &  0.139 &  &  0.115   &   0.154 &  \\
                \midrule
                Sample size & 202 & 218 & & 144 & 166 & & 96 & 130 &\\
				\bottomrule
			\end{tabular}
            \begin{tablenotes}
            \scriptsize
            \item \emph{Note:} Sample averages and p-values of mean-comparison tests; Reference groups: Age: $17 - 19$; Caregiver: no caregiver; Federal State: Other; Female: Male; Migration Parents: Both foreign; Mother tongue German: mother tongue not German; Work (in hours per week): None.
            \end{tablenotes}
            \end{threeparttable}
		\end{table}
	\end{landscape}

	\begin{landscape}
		\begin{table}[!htbp]
        \begin{threeparttable}
			\centering
			
			\footnotesize
			\caption{Learning Preconditions}
            \label{tab:learningpreconditions}
			\begin{tabular}{lccccccccc}
				\toprule
				& \multicolumn{3}{c}{Sample A} 
				& \multicolumn{3}{c}{Sample B} 
				& \multicolumn{3}{c}{Sample C} \\
				\cmidrule(lr){2-4} \cmidrule(lr){5-7} \cmidrule(lr){8-10}
				Variable
				& Cohort 1 & Cohort 2 & p-value
				& Cohort 1 & Cohort 2 & p-value
				& Cohort 1 & Cohort 2 & p-value \\
				\midrule
				Abitur   &  0.970 & 0.991   & 0.125  & 0.972 & 0.988  & 0.318 &  0.958   &  0.985 & 0.226  \\
				Advanced Math Course   &   0.351   &   0.390   &  0.417  &   0.319   &  0.416 & 0.081   &  0.292 &  0.408 &  0.073 \\
				%	\midrule
				Business Administration   &  0.584  &    0.633   &   0.306  &  0.660  &    0.687 & 0.614 &  0.740   &  0.654 & 0.170  \\
				%    \midrule
				\underline{High-School Type}  &  &  &  0.049  &  & & 0.059  &  & & 0.088 \\
				\hspace{.1cm}  Foreign  &  0.030   & 0.009   &   &  0.028  &            0.012 &  &  0.031  &  0.015 &  \\
				\hspace{.1cm}  Gesamtschule &  0.054  &   0.106 &   &  0.049    &      0.127  &  &   0.042   &   0.138 &  \\
				\hspace{.1cm} Gymnasium   &  0.787 &  0.803  &   &  0.792 & 0.771 &  &  0.844  &   0.785 &  \\
				%	\midrule
				\underline{Last Math Grade}  &  &  &  0.106 &  & &  0.030   &  & & 0.496  \\
				\hspace{.1cm}  Very Good  &  0.183     &        0.106  &   &  0.208   &   0.096 &  &   0.177   & 0.123 &  \\
				\hspace{.1cm} Good  &  0.371   &  0.427  &   &   0.368 &  0.470 &  &  0.479   &     0.446 &  \\
				\hspace{.1cm} Satisfactory  &  0.332  &     0.321    &   &  0.312    &     0.295  &  &  0.240    & 0.285 &  \\
				%	\midrule
				\underline{Outcome Expected}  &  &  & 0.966   &  & & 0.813 &  & & 0.776  \\
				\hspace{.1cm} Very Good  &  0.050    &     0.060  &   &  0.028     &     0.048 &  &  0.031   &  0.046 &  \\
				\hspace{.1cm} Good   &  0.490  &      0.495  &   &  0.507  &            0.512 &  &   0.542 &  0.492 &  \\
				\hspace{.1cm} Satisfactory   &  0.411  &  0.399  &   &  0.424   &   0.398 &  &  0.375  &  0.423 &  \\
				\underline{Outcome Goal} &  &  & 0.922   &  & & 0.718   &  & & 0.697  \\
				\hspace{.1cm} Very Good  &   0.347  &      0.349    &   &  0.312    &    0.349  &  &  0.312  &  0.323 &  \\
				\hspace{.1cm} Good  &  0.589  &  0.596  &   &  0.611 &   0.590  &  &  0.615  &   0.631 &  \\
				%	    \midrule
				\underline{Previous Math Course} &  &  & 0.906   &  & & 0.925  &  & & 0.815 \\
				\hspace{.1cm} One  &  0.059  &  0.069  &   &  0.056   &           0.060   &  & 0.062   &   0.085 &  \\
				\hspace{.1cm}  Multiple  &  0.064  &    0.069   &   &  0.062   &   0.072  &  &  0.073  &    0.077 &  \\
				%	    \midrule
				At least One Previous Statistics Course  & 0.059 & 0.073 & 0.567  & 0.062 & 0.072 & 0.733 & 0.052 & 0.077 & 0.461 \\
				Reversed High-School Grade &  2.716 &    2.675   & 0.489  &  2.792    &   2.757 & 0.605  &  2.908   &    2.745 &  0.029 \\
				Statistical Software Experience   & 0.040  &     0.046   &  0.752  &  0.021  &    0.048   &  0.195 &  0.031    &       0.038 & 0.773 \\
				Study Experience   &  0.035  &     0.060   &  0.231   &  0.035  &    0.054 &  0.411 &   0.031   &      0.062 &  0.298 \\
				Vocational Training   &  0.173  &     0.110  & 0.063  &  0.194  &   0.096  &  0.014 &  0.167 &    0.115 & 0.270  \\
				 \midrule
                Sample size & 202 & 218 & & 144 & 166 & & 96 & 130 &\\
				\bottomrule
			\end{tabular}
            \begin{tablenotes}
            \scriptsize
            \item \emph{Note:} 
            Sample averages and p-values of mean-comparison tests; Reference groups: Abitur: Other; Advanced Math Course: None; Business Administration: Economics; High-School Type: Other; Last Math Grade: Pass or fail; Outcome Expected: Pass or fail; Outcome Goal: Satisfactory, pass, or fail; Previous Math Course: None; At least One Previous Statistics Course: None; Statistical Software Experience: None; Study Experience: None; Vocational Training: None.
            \end{tablenotes}
        \end{threeparttable}
		\end{table}
	\end{landscape}

	\begin{landscape}
		\begin{table}[!htbp]
            \begin{threeparttable}
			\centering
			\footnotesize
			\caption{Likert Data -- First Questionnaire}
            \label{tab:likertq1}
			\begin{tabular}{lccccccccc}
				\toprule
				& \multicolumn{3}{c}{Sample A} 
				& \multicolumn{3}{c}{Sample B} 
				& \multicolumn{3}{c}{Sample C} \\
				\cmidrule(lr){2-4} \cmidrule(lr){5-7} \cmidrule(lr){8-10}
				Variable
				& Cohort 1 & Cohort 2 & p-value
				& Cohort 1 & Cohort 2 & p-value
				& Cohort 1 & Cohort 2 & p-value \\
				\midrule
				\underline{Perception of Statistics} &  &  &  &  & &  &  & &  \\
				\hspace{.1cm} Affection  & 0.819  & 0.742 & 0.458 & 0.705   &  0.709 &  0.974  & 0.762 & 0.678 &  0.541 \\
				\hspace{.1cm}  Difficulty  & 1.031   &          1.017 &  0.868 &             1.078        &      1.050    &         0.762 & 1.073  &           1.081 & 0.944\\
				\hspace{.1cm}  Effort & 1.787  &  1.852 &  0.449   & 1.810  & 1.852 & 0.659 & 1.837 & 1.887 & 0.658 \\
				\hspace{.1cm} Extrinsic Motivation & 1.611   & 1.650  & 0.702  & 1.521 &  1.633 & 0.344 &  1.641     &        1.583 & 0.684  \\
				\hspace{.1cm}  Interest &  1.282 &  1.382 & 0.347  & 1.222  & 1.348 &  0.310  & 1.344     & 1.329 & 0.917 \\
				\hspace{.1cm} Self-Concept & 1.210   & 1.142  & 0.433 & 1.124  & 1.117   &  0.945 & 1.192 &  1.069 &  0.311 \\
				\hspace{.1cm}  Value & 1.658 &  1.890 & 0.017   &  1.653 & 1.894 & 0.036 & 1.681 & 1.872 &  0.167 \\
				% \midrule
				\underline{Learning} &  &  &  &  & &  &  & &  \\                                                        
				\hspace{.1cm} Critical Thinking & 0.727 & 0.853      & 0.179  & 0.736   &  0.830 &  0.390    &  0.831 &   0.843 & 0.925  \\
				\hspace{.1cm} Environment & 1.411 & 1.383 & 0.780 & 1.474 & 1.393 & 0.467 &  1.504 & 1.403  &  0.425 \\
				\hspace{.1cm}  Procrastination &  -0.653   & -0.783 & 0.279 & -0.758 & -0.834 &  0.584  & -0.796  & -0.812 &  0.919  \\
				\hspace{.1cm} Repetition  & 1.432  & 1.420 & 0.892    & 1.403   &  1.446   & 0.684 &  1.495   &  1.383 &  0.346 \\
                \hspace{.1cm} Peer  & 0.515  & 0.792 & 0.051    & 0.551   &  0.659   & 0.508 &  0.601   &  0.754 &  0.423 \\
				\hspace{.1cm}  Self-Regulation & 1.683 &  1.705  & 0.796   &  1.724 &  1.722  & 0.983  &  1.748 &  1.669 & 0.485 \\
				\hspace{.1cm}  Test-Anxiety &  -0.717      & -0.570  & 0.297  &  -0.789 &  -0.523 & 0.104  &  -0.713 & -0.609 & 0.595 \\
			 \midrule
                Sample size & 202 & 218 & & 144 & 166 & & 96 & 130 &\\
				\bottomrule
			\end{tabular}
            \begin{tablenotes}
            \scriptsize
            \item \emph{Note:} Sample averages and p-values of mean-comparison tests; 7-point Likert scale from -3 (totally disagree) over 0 (neutral) to 3 (totally agree); In the following the upward arrow $\uparrow$ abbreviates "higher value means"; \underline{Perception of statistics:} Affection: $\uparrow$ more positive attitude towards statistics; Difficulty: $\uparrow$ considers statistics more difficult; Effort: $\uparrow$ tries to work harder in the course; Extrinsic motivation: $\uparrow$ more ambitious to achieve a good grade in the course;  Interest: $\uparrow$ more interested in statistics; Self-concept: $\uparrow$ more optimistic to master the subject; Value: $\uparrow$ values statistics more for further study, job, or daily life. \underline{Learning:} Critical thinking: $\uparrow$ more able to critically reflect course material; Environment: $\uparrow$ more organized to plan self-learning; Procrastination: $\uparrow$ procrastinates longer when learning; Repetition: $\uparrow$ uses a more repetitive learning style; Peer: $\uparrow$ engages more in collaborative learning activities with peers; Self-regulation: $\uparrow$ monitors and adapts learning progress more; Test-anxiety: $\uparrow$ fears the exam more. 
            \end{tablenotes}
        \end{threeparttable}
		\end{table}
	\end{landscape}

	\begin{table}[!htbp]
        \begin{threeparttable}
		\centering
		\footnotesize
		\caption{Outcome variables}
        \label{tab:outcomes}
		\begin{tabular}{llccc}
			\toprule
			&Variable & Cohort 1 & Cohort 2 & p-value \\
			\midrule
			\textit{Sample A}&Exam Points &  28.146 &  27.456  &  0.517 \\
			($n_{C1} = 202$ / $n_{C2} = 218$)&Passed Exam &  0.802  &  0.743  & 0.152 \\
			\midrule
			\textit{Sample B}&Knowledge Retention Points &   5.521   & 5.380  &  0.553\\
            ($n_{C1} = 144$ / $n_{C2} = 166$)&&&&\\
			\midrule
			\textit{Sample C}&\underline{Perception of Statistics} &  &  &  \\
			($n_{C1} = 96$ / $n_{C2} = 130$)&\hspace{.1cm} Affection         &  0.783 &   0.950  & 0.225 \\
			&\hspace{.1cm} Difficulty       &   0.516 &   0.471 &   0.705\\
			&\hspace{.1cm} Effort            &  0.139  &  0.333  & 0.236 \\
			&\hspace{.1cm} Extrinsic Motivation  &  1.060  &  0.954  & 0.510 \\
			&\hspace{.1cm} Interest          &  0.812  &  0.660  & 0.306 \\
			&\hspace{.1cm} Self-Concept      &  1.033 &    1.314  & 0.033 \\
			&\hspace{.1cm} Value            &  1.238  &  1.288  &  0.724 \\
			&\underline{Learning}              &  &  &  \\
			&\hspace{.1cm} Critical Thinking &  -0.335 &  -0.357 & 0.889  \\
			&\hspace{.1cm} Environment       & 0.973  &  1.052  &  0.546\\	
			&\hspace{.1cm} Procrastination   & 0.042 &  -0.334  & 0.019  \\
			&\hspace{.1cm} Peer        &  0.038 &   -0.341  & 0.067 \\
			&\hspace{.1cm} Self-Regulation   &   1.050  &  1.120 &  0.558 \\
			&\hspace{.1cm} Test-Anxiety      &  -0.669  & -0.885 & 0.280 \\
			&\underline{Course Perception}     &  &  &  \\
			&\hspace{.1cm} Boredom     & -0.753   & -0.733 & 0.919  \\
			&\hspace{.1cm} Enjoyment  &   0.542   & 0.174  &  0.016 \\
			&\hspace{.1cm} Frustration &   -1.794  & -1.908  & 0.462 \\
			&\hspace{.1cm} Self-Concept &  0.606  &  0.959  &  0.031\\
			\bottomrule
		\end{tabular}
        \begin{tablenotes}
        \scriptsize
        \item \emph{Note:} Sample averages and p-values of mean-comparison tests; 7-point Likert scale from -3 (totally disagree) over 0 (neutral) to 3 (totally agree); In the following the upward arrow $\uparrow$ abbreviates "higher value means"; \underline{Perception of statistics:} Affection: $\uparrow$ more positive attitude towards statistics; Difficulty: $\uparrow$ considers statistics more difficult; Effort: $\uparrow$ tries to work harder in the course; Extrinsic motivation: $\uparrow$ more ambitious to achieve a good grade in the course;  Interest: $\uparrow$ more interested in statistics; Self-concept: $\uparrow$ more optimistic to master the subject; Value: $\uparrow$ values statistics more for further study, job, or daily life. \underline{Learning:} Critical thinking: $\uparrow$ more able to critically reflect course material; Environment: $\uparrow$ more organized to plan self-learning; Procrastination: $\uparrow$ procrastinates longer when learning; Peer: $\uparrow$ engages more in collaborative learning activities with peers; Self-regulation: $\uparrow$ monitors and adapts learning progress more; Test-anxiety: $\uparrow$ fears the exam more. \underline{Course perception:} Boredom: $\uparrow$ perceives classroom sessions as more boring; Enjoyment: $\uparrow$ enjoys classroom sessions more; Frustration: $\uparrow$ gets more frustrated in classroom sessions; Self-concept: $\uparrow$ more optimistic to master the subject during classroom sessions.
        \end{tablenotes}
        \end{threeparttable}
	\end{table}

\section{Methodology} \label{s4_meth}

\subsection{Causal Structure}\label{s41_cg}

We aim to determine the effect of changing the instructional format from lecture-based teaching to flipped classroom on several student outcomes: exam performance, knowledge retention, learning strategies, perception of statistics, and course perception. Our cohort design involves measuring these outcomes at the end of the course. A potential concern with such a design is selection bias based on covariates, even if we expect subsequent cohorts to be generally similar. For example, if certain student characteristics predict dropout in the flipped classroom more than in the blended format, the resulting observed samples would be systematically different. Although the mean comparisons presented in Section \ref{s3_data} suggest balance for most additional covariates, this is insufficient to ensure balance across the entire joint covariate distribution. Consequently, our analysis includes additional covariates related to sociodemographics, learning preconditions, learning strategies, and perception of statistics to mitigate this potential bias.

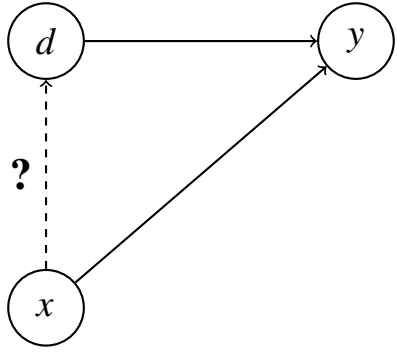
\begin{figure}[!htbp]
    \caption{Causal diagram}
    \label{fig:dag}
    \centering
    \footnotesize
    \begin{tikzpicture}[
        node distance=1.25cm, 
        every node/.style={draw, circle, minimum size=1cm, font=\large}, 
        every path/.style={thick,->},
    ]
        % Nodes
        \node (d) at (-.7,0)  {$d$};               
        \node (y) at (3.4,0) {$y$};    
        \node (x) [below=of d, yshift = -1.25 cm] {$x$};  
        
        % Edges 
        \path (d) edge (y); 
        \path (x) edge (y);  
        \path[draw, thick, dashed] (x) -- node[left=-5.2pt, draw=none] {\scalebox{1.15}{\textbf{?}}} (d);  

        % Add text labels next to nodes
        \node (F) [draw=none, left=0.25cm of d] {};    
        \node[draw=none, right=0.25cm of y] {};
        \node[draw=none, left=0.25 cm of x, yshift=-.25 cm] (C) {};   
        
       \renewcommand{\labelitemi}{\raisebox{0.025ex}{\scalebox{0.6}{$\bullet$}}} 
    \end{tikzpicture}
\end{figure}

Figure \ref{fig:dag} visualizes the causal structure of our study. Let $y$ be an outcome variable, $d$ be an indicator taking a value of one if part of the flipped cohort and zero otherwise, and $x$ be a set of additional covariates. Domain knowledge suggests a direct effect of $x$ on $y$. However, as argued in the previous paragraph, we acknowledge the possibility that $x$ also affects $d$. This dual role of $x$ needs careful consideration in our analysis.

\subsection{Estimation Approaches}\label{s42_a}

We consider the following general model specification implied by Figure \ref{fig:dag},
\begin{equation}
    \label{eq:model}
    y_{i} = g(d_{i}, x_{i}) + u_{i} \, , \quad d_{i} = m(x_{i}) + v_{i} \, , \quad i = 1, \ldots, n \, ,
\end{equation}
where $u_{i}$ is an additive error term. A key feature of our model is that the additional covariates, $x_{i}$, are allowed to be related to $d_{i}$ through the propensity score $m(x_{i})$. Simultaneously, the outcome $y_{i}$ can be affected by both $d_{i}$ and $x_{i}$ through the function $g(d_{i}, x_{i})$.

We are interested in the average treatment effect (ATE),
\begin{equation*}
    \theta = \mathbb{E}[g(1, x_{i}) - g(0, x_{i})] \, .
\end{equation*}
The ATE can be identified under standard causal inference assumptions:
\begin{enumerate}[i)]
    \item \textit{(unconfoundedness)} given $x_{i}$, $d_{i}$ is independent of the potential outcome.
    \item \textit{(common support)} $0 < \text{Pr}(d_{i} = 1 \mid x_{i} = x) < 1$ for every $x$ in the range of $x_{i}$.
    \item \textit{(stable unit treatment value)} $y_{i} = y_{1i} d_{i} + y_{0i} (1 - d_{i})$, where $y_{1i}$ and $y_{0i}$ are the potential outcomes of the unit $i$ when part of the flipped and blended cohort, respectively.
\end{enumerate}
Although we are mainly interested in the ATE, it is important to note that our most general specification does not restrict the treatment effects to be homogeneous across students.

The classical literature on estimating ATEs introduced several methods, including those based on covariate adjustment, propensity score adjustment, or doubly robust methods (which combine both approaches), as proposed in the work of \textcites{robins, hahn, hirano, abadie}. A significant limitation of these methods is that they are developed for low-dimensional settings, meaning that they are suited for settings with only few additional covariates, which is not compatible with the large number of covariates in our data. As a consequence, our study employs the double/debiased machine learning (DML) approach proposed by \textcite{chernozhukov}. DML complements the classical literature by allowing for valid inferential statements about a low-dimensional parameter of interest, while integrating advanced machine learning (ML) techniques that are known for their strong predictive capabilities. In our study, we use the doubly robust estimator for the ATE as described in Section 5.1 of \textcite{chernozhukov}.

An important assumption of the DML approach to obtain $\sqrt{n}$ rate of convergence and asymptotic normality is that the ML technique used should estimate its parameters at a rate of $o(n^{-1/4})$. This assumption is reasonable for many ML methods under additional structural assumptions on its parameters, such as sparsity. In our setting, with a large number of very correlated potential confounders relative to the sample size of the data, ML techniques that use regularization/penalization or other types of dimension reduction techniques, e.g., lasso (Lasso) or ridge regression (Ridge), best subset selection (BSS), as well as methods that used derived input directions, e.g., principal component regression (PCR) or partial least squares (PLS) are particularly useful. Importantly, the convergence rates of these ML methods in a high-dimensional setting are well understood. See, among others, \textcites{bickel, buhlmann, belloni2012, belloni2014} for lasso regression, \textcite{audibert} for ridge regression, \textcite{kozbur} for forward selection, \textcite{green} for PCR, and \textcite{cook} for PLS.

Recall that our data set includes numerous additional covariates, which are highly correlated. Among these, most of the covariates are Likert items intended to measure various psychological constructs, including, for example, procrastination tendency. The procrastination scale proposed by \textcite{patzelt} consists of 19 questions, that is, 19 distinct measures of procrastination tendencies. Importantly, these measures are expected to be related to a lower-dimensional set of latent variables. To provide a more specific example, let $Z$ be a matrix with dimensions $n \times 19$, which contains the 19 measures of procrastination for each of the $n$ units. Then, $Z$ can be represented as a factor model,
\begin{equation*}
    Z = F \Lambda^{\prime} + E \, ,
\end{equation*}
where $\Lambda$ is a $p \times k$ matrix of factor loadings, $F$ is a $n \times k$ matrix of factors, $k < 19$, and $E$ is a $n \times p$ error matrix. Using exploratory factor analysis, \textcite{helmke} identified three distinct factors, which they interpreted as i) core issues of procrastination, ii) lack of foresight, and iii) being late.\footnote{In particular, 13 questions measure i) core issues of procrastination, four questions measure ii) lack of foresight, and 2 questions measure iii) being late.} Therefore, beyond the challenge posed by numerous highly correlated covariates, it is essential to take into account that most of our variables might have a lower-dimensional representation, which justifies our selection of machine learning techniques.

Furthermore, we want to highlight that there is a close connection between some of the ML techniques that we consider and factor analysis. For example, \textcite{chamberlain} showed that the principal component analysis (PCA), which is the first step of PCR, is a valid estimator for approximate factor models. In contrast to factor analysis, where factors are typically estimated by maximum likelihood, PCA can be interpreted as a least squares approach that avoids a strong distributional assumption about the error matrix. The drawback is that $k$ needs to be large enough. In addition, Ridge and PCR are closely related. To see this, let $X$ denote some $n \times p$ design matrix and let $y$ be some outcome variable of dimension $n \times 1$. Then, by the singular value decomposition, we have $X = U \Sigma V^\prime$, where $U$ is an orthogonal matrix of dimension $n \times p$ with columns that are a basis of the column space of $X$, $V$ is an orthogonal matrix of dimension $p \times p$ with columns that are a basis of the row space of $X$, and $\Sigma$ is a diagonal matrix of dimension $p \times p$ with diagonal entries $\sigma_{1} \geq \sigma_{2} \geq \ldots \geq \sigma_{p}$. Hence, the coefficients of a PCR of $y$ on $X$ can be obtained as
\begin{equation*}
	\hat{\beta}_{\text{PCR}}(k) = \sum_{j = 1}^{k} u_{j} u_{j}^{\prime} y \, ,
\end{equation*} 
where $u_{j}$ denotes the $j$-th column of $U$. Likewise,
\begin{equation*}
	\hat{\beta}_{\text{Ridge}}(\lambda) = \sum_{j = 1}^{p} \bigg(\frac{\sigma_{j}^{2}}{\sigma_{j}^{2} + \lambda}\bigg) u_{j} u_{j}^{\prime} y 
\end{equation*}
gives us the corresponding coefficients of Ridge. Hence, while PCR truncates the coordinates of $y$ with respect to the basis $U$, that is, computes the coefficients only based on the first $k < p$ coordinates corresponding to the $k$ largest singular values, Ridge computes the coefficients based on all coordinates but with decaying weights.

\textcite{frank} extensively compared the predictive performance of BSS, PCR, PLS, and Ridge against OLS in various scenarios, including those with highly correlated covariates. Their results indicate that all dimension reduction techniques significantly outperform OLS, with Ridge being their favorite for minimizing prediction error. They proposed the following performance ranking: Ridge > PLS > PCR > BSS > OLS. Although the differences between Ridge, PLS, and PCR were often marginal, they attributed Ridge's generally better performance to its smoother shrinkage behavior compared to the more discrete and extreme behavior of the other methods. \textcite[Chapter 3.6]{hastie} provide an additional comparison, adding Lasso to the list of ML techniques and placing its performance between Ridge and BSS.

\section{Results}\label{s5_res}

Based on the identification strategy outlined in the previous section, we analyze the effects of changing the instructional format from lecture-based teaching to a flipped classroom on various student outcomes. To mitigate potential biases, such as those induced by systematic differences in dropout rates, we adjusted for a rich set of covariates in our analysis. Specifically, we account for all covariates detailed in Section \ref{s3_data}, which encompasses sociodemographics, learning preconditions, initial perceptions of statistics, and initial learning strategies. Our comprehensive set of 103 variables includes 67 Likert items related to 14 distinct scales. Consequently, in our smallest sample (Sample C), the number of observations is barely twice as large as the number of covariates. The dimensionality of this problem underscores the necessity of employing appropriate estimation techniques capable of providing valid inference in this high-dimensional setting.

For our analysis, we employ the Double Machine Learning (DML) approach of \textcite{chernozhukov}, utilizing ridge regression to estimate $g(d_{i}, x_{i})$ and ridge classification for $m(x_{i})$. As argued in the previous section, estimation techniques based on penalization, such as ridge regression, are particularly useful for our setting with highly correlated covariates. In addition, to highlight the necessity of DML in our high-dimensional setting, we also present results from a standard regression adjustment,
\begin{equation*}
    y_{i} = \theta \, d_{i} + x_{i}^{\prime} \beta + u_{i} \, .
\end{equation*}
For this classical approach, we consider three covariate sets: (1) the full set (also used in DML), (2) a reduced set using the means of 13 out of 14 Likert scales (replacing 63 distinct items), and (3) another set reducing the dimension of 63 distinct items (associated with 13 out of 14 Likert scales) to the first 13 principal components (see Online Appendix \ref{app:pca}). We do not reduce the dimensionality of the Likert scale for repetition. Our prior item analysis (see Online Appendix \ref{app:item_analysis}) indicates that the repetition scale cannot be adequately represented by a unidimensional latent construct. Hence, we include the four items without dimension reduction in OLS (2) and OLS (3). For the DML results, we present two specifications: the general form outlined in Section \ref{s42_a}, and a restricted specification in which we assume $g(d_{i}, x_{i}) = \theta \, d_{i} + h(x_{i})$ and estimate $h(x_{i})$ by ridge regression.\footnote{All our analyses were done using the statistical software R \parencite{r_software} and the following packages: \textit{DoubleML} \parencite{dml_software} (for DML), \textit{mlr3} \parencite{mlr3_software} (for ML techniques within \textit{DoubleML}), and \textit{glmnet} \parencite{glmnet_software} (for ridge regression and classification).} Table \ref{tab:estimated_ates} summarizes our results.

\begin{table}[!htbp]
	\centering
	\footnotesize
    \begin{threeparttable}
	\caption{Estimated Average Treatment Effects (ATEs) for Different Outcomes}
    \label{tab:estimated_ates}
	\begin{tabular}{llccccc}
		\toprule
		&Outcome variable & \multicolumn{5}{c}{Estimated ATE} \\
        \cmidrule(lr){3-7}
        && OLS (1) & OLS (2) & OLS (3) & DML (1) & DML (2)\\
		\midrule
		\textit{Sample A} &Exam Points & -0.056 &  0.262 &  0.322 & -0.416 & -0.634 \\ 
        ($n_{C1} = 202$ / $n_{C2} = 218$)&& (1.035) & (0.929) & (0.944) & (0.917) & (0.956) \\ 
		&Passed Exam & -0.053 & -0.040 & -0.037 & -0.046 & -0.053 \\ 
        && (0.043) & (0.040) & (0.040) & (0.038) & (0.039) \\ 
		\midrule
		\textit{Sample B} &Knowledge Retention Points & -0.058 &  0.029 &  0.050 & -0.146 & -0.164 \\ 
        ($n_{C1} = 144$ / $n_{C2} = 166$)&& (0.276) & (0.241) & (0.243) & (0.227) & (0.237) \\ 
		\midrule
		\textit{Sample C}&\underline{Perception of Statistics} &  &  & &&  \\[0.5em]
		($n_{C1} = 96$ / $n_{C2} = 130$)&\hspace{.1cm} Affection        &  0.345$^{*}$ &  0.196 &  0.157 &  0.230$^{*}$ &  0.214 \\ 
        && (0.200) & (0.155) & (0.156) & (0.126) & (0.134) \\ 
        &\hspace{.1cm} Difficulty        &  0.056 &  0.017 & -0.012 & -0.047 & -0.050 \\ 
        && (0.148) & (0.128) & (0.118) & (0.111) & (0.117) \\ 
        &\hspace{.1cm} Effort            &  0.316 &  0.370$^{*}$ &  0.302 &  0.208 &  0.203 \\ 
        && (0.223) & (0.200) & (0.194) & (0.161) & (0.166) \\ 
        &\hspace{.1cm} Extrinsic Motivation  & -0.116 & -0.142 & -0.226 & -0.085 & -0.088 \\ 
        && (0.219) & (0.185) & (0.190) & (0.152) & (0.162) \\ 
        &\hspace{.1cm} Interest          & -0.166 & -0.165 & -0.245 & -0.163 & -0.165 \\ 
        && (0.206) & (0.172) & (0.160) & (0.146) & (0.152) \\ 
		&\hspace{.1cm} Self-Concept      &  0.362$^{*}$ &  0.303$^{**}$ &  0.277$^{*}$ &  0.300$^{**}$ &  0.293$^{**}$ \\ 
        && (0.206) & (0.150) & (0.157) & (0.127) & (0.133) \\ 
		&\hspace{.1cm} Value             & -0.166 & -0.177 & -0.195 & -0.021 &  0.005 \\ 
        && (0.202) & (0.167) & (0.159) & (0.139) & (0.148) \\ 
		&\underline{Learning}              & & & & & \\[0.5em]
        &\hspace{.1cm} Critical Thinking &  0.073 &  0.013 & -0.022 & -0.069 & -0.059 \\ 
        && (0.203) & (0.164) & (0.165) & (0.148) & (0.156) \\ 
		&\hspace{.1cm} Environment       &  0.075 &  0.061 &  0.025 &  0.090 &  0.089 \\
        && (0.157) & (0.143) & (0.135) & (0.117) & (0.126) \\ 
		&\hspace{.1cm} Procrastination   & -0.225 & -0.368$^{**}$ & -0.294$^{*}$ & -0.344$^{**}$ & -0.365$^{**}$ \\ 
        && (0.195) & (0.165) & (0.157) & (0.150) & (0.158) \\ 
        &\hspace{.1cm} Peer        & -0.276 & -0.399$^{*}$ & -0.403$^{*}$ & -0.402$^{**}$ & -0.406$^{**}$ \\ 
        && (0.307) & (0.215) & (0.228) & (0.195) & (0.207) \\ 
        &\hspace{.1cm} Self-Regulation   &  0.207 &  0.172 &  0.192 &  0.115 &  0.096 \\ 
        && (0.155) & (0.128) & (0.125) & (0.113) & (0.117) \\ 
		&\hspace{.1cm} Test-Anxiety      & -0.178 & -0.212 & -0.202 & -0.272 & -0.271 \\
        && (0.260) & (0.178) & (0.175) & (0.171) & (0.189) \\ 
		&\underline{Course Perception}   & & & & & \\[0.5em]
        &\hspace{.1cm} Boredom   &  0.297 &  0.010 & -0.037 & -0.003 &  0.001 \\ 
        && (0.253) & (0.230) & (0.230) & (0.193) & (0.201) \\ 
		&\hspace{.1cm} Enjoyment  & -0.503$^{**}$ & -0.368$^{**}$ & -0.367$^{**}$ & -0.362$^{**}$ & -0.371$^{**}$ \\ 
        && (0.203) & (0.181) & (0.182) & (0.146) & (0.153) \\ 
        &\hspace{.1cm} Frustration & -0.106 & -0.163 & -0.192 & -0.139 & -0.137 \\ 
        && (0.237) & (0.180) & (0.184) & (0.153) & (0.160) \\ 
		&\hspace{.1cm} Self-Concept &  0.202 &  0.374$^{*}$ &  0.367$^{*}$ &  0.370$^{**}$ &  0.381$^{**}$ \\ 
        && (0.231) & (0.193) & (0.201) & (0.161) & (0.168) \\ 
		\bottomrule
	\end{tabular}
    \begin{tablenotes}
        \scriptsize
        \item \emph{Note:} Robust standard errors in parentheses; DML uses ridge regression/classification to estimate $g(d_{i}, x_{i})$ and $m(x_{i})$ in \eqref{eq:model}; tuning parameters for ridge regression/classification are determined by ten-fold cross-validation; DML results are based on five-fold cross-fitting and 100 repetitions.
        \item $^{*}$p$<$0.1; $^{**}$p$<$0.05; $^{***}$p$<$0.01
    \end{tablenotes}
    \end{threeparttable}
\end{table}

We find no significant effects of the change in instructional format on exam points. Beyond statistical insignificance, the estimated effect sizes are not very large. To illustrate, the estimate of DML (2), which shows the largest magnitude, suggests that students in the flipped classroom cohort scored on average 0.6 points lower, with a standard error of larger magnitude. In line with our expectations, the reduction in dimensionality employed in OLS (2) and (3) improves the precision of the estimates. Furthermore, all estimated effects on passing the exam are insignificant as well. However, the estimated effects are much closer to each other than for the exam points. Here, DML (2) predicts that students in the flipped cohort have on average a roughly five percentage points lower probability of passing the exam. Similarly to exam performance, we do not find any significant effects on knowledge retention with estimated effect sizes being quite small.

Our analysis of end-of-course perception of statistics shows that the majority of associated psychological constructs exhibit insignificant effects, regardless of the estimation approach. An exception is self-concept, which is significant at the 5\% level for both DML approaches. Notably, our two DML estimates for this outcome are very close and suggest that altering the instructional format significantly increases students' perceived competence in mastering statistics by an average of 0.3 points on the 7-point Likert scale. Moreover, we find considerable positive effects on affection and effort, although only some of these are statistically significant at the 10\% level. Similarly, our analysis of end-of-course learning strategies reveals insignificant effects for most associated psychological constructs, with procrastination and peer being notable exceptions. For procrastination, we find significant negative effects for all estimation methods except OLS (1). Although estimated reductions in procrastination are comparable between OLS (3) and both DML approaches, DML estimates exhibit greater precision. Our findings indicate that the change in instructional format significantly reduces students' tendency to procrastinate by an average of roughly 0.35 points on the 7-point Likert scale. Likewise, for peer, we find significant negative effects for all estimation methods except OLS (1) with DML estimates again showing greater precision. The instructional format change significantly reduces students' engagement in collaborative learning activities with peers by approximately 0.4 points. Furthermore, we observe substantial negative effects on test anxiety, although these are not statistically significant.

Lastly, our analysis reveals considerable and significant effects of the change in instructional format on course perception. Specifically, our OLS (2), OLS (3), and DML estimates indicate a positive effect on self-concept, suggesting that altering the format significantly increases students' perceived competence in mastering the course by approximately 0.35 points on the 7-point Likert scale. Although OLS (2) and OLS (3) yield similar point estimates for self-concept, they lack the necessary precision. Furthermore, we find significant negative effects on course enjoyment, regardless of the estimation method used. Again, OLS (2) and OLS (3) show point estimates similar to DML, but DML estimates are more precise. Our findings suggest that the instructional format change significantly reduces students' enjoyment of the course by roughly 0.4 points on the 7-point Likert scale. For all other outcomes related to course perception, we find no significant effects.

When comparing estimation approaches, we observe that most of the time the OLS estimates align in the direction with the DML, and some estimates even exhibit comparable magnitudes. However, as expected in our high-dimensional setting, OLS lacks the robustness and precision of DML. Furthermore, the estimates of DML (1) and DML (2) are remarkably similar. Given that DML (1) assumes homogeneous student-specific treatment effects while DML (2) does not, this similarity suggests that the effects of shifting from lecture-based teaching to a flipped classroom are relatively similar across students. Hence, imposing treatment effect homogeneity across students might be reasonable in our setting.

Our findings are in line with learning-theoretical expectations. Flipped classrooms provide students with active self-testing opportunities and feedback mechanisms throughout weekly learning phases and classroom sessions, as noted by \textcite{o_flaherty}. Combining self-testing with timely and relevant feedback is theoretically and empirically supported to improve learning (see, for example, \textcites{butler2007, mcdaniel, schwerter}). Consequently, as pointed out by \textcites{bacon, roediger_incoll}, students in the flipped classroom are more likely to be aware of their proficiency and better equipped to compare their performance with peers during classroom sessions than those in the lecture-based teaching setting. Therefore, it is plausible that changing the instructional format to a flipped classroom positively affects students' self-perceived capabilities in mastering the subject and the classroom session. Furthermore, estimated negative effects on procrastination indicate that students in the flipped classroom cohort were less prone to procrastinate (or at least perceived themselves as such), a desirable (or problematic)\footnote{Students that perceive themselves as better prepared, although not engaging ideally with the learning material, might see no need to further engage with the content or adapt their learning strategy.} outcome. As argued by \textcites{ericson, clark2019}, flipped classrooms offer structured weekly schedules that foster spaced learning, a well-established strategy to reduce procrastination.\footnote{Spaced learning, in contrast to learning in condensed blocks, involves distributing learning activities in regularly spaced intervals over time (see, for example, \textcites{rodriguez_spacing, schwerter}).} The estimated decrease in course enjoyment can probably be attributed to the increased active involvement required by flipped classroom sessions, which students may perceive as more demanding. This is consistent with \textcite{akccayir}, who find that students associate flipped curricula with an increased workload. In addition, the enhanced recognition of errors and misconceptions among students in the flipped cohort during interactive classroom activities such as clicker polls and discussions, as opposed to the passive learning environment of the lecture-based cohort, could also have contributed to a less favorable perception of the flipped classroom format. In this context, it should also be mentioned that \textcite{van_alten} did not find strong evidence that students are more satisfied with flipped learning environments. Finally, our flipped classroom was designed to encourage collaborative learning during interactive classroom activities. However, we observe reduced student engagement in peer-based learning activities. As a consequence, students in the flipped cohort missed out on the full benefits of the intended collaborative learning environment, which was designed to improve critical thinking and problem solving skills \parencite{g2012}.

Since there are contradictory recommendations in the psychometric literature on handling Likert scales as dependent variables, we provide additional results for Sample C in Online Appendix \ref{app:irt_results}. In particular, we utilize Item Response Theory (IRT) models to estimate the underlying latent constructs, as an alternative to the unit averages associated with Classical Test Theory (CTT).\footnote{IRT traits were estimated using the statistical software R \parencite{r_software} and the \textit{mirt} package \parencite{mirt_software}.} A key benefit of the IRT approach is that its estimates are on an interval scale, which allows a more meaningful interpretation of the underlying psychometric construct (see, among others, \textcites{jabrayilov}{harwell}). However, we opted for the CTT-based approach in our main analysis, given that empirical studies frequently report only minor differences between the two approaches, especially when the scales are well constructed and unidimensional (see, among others, \textcites{xu}{sebille}), and to avoid the additional distributional assumptions inherent in IRT models in favor of a more robust approach. In addition, \textcite{sijtsma2024recognize} recently show that unit averages provide a mathematically justified ordinal measure of the underlying latent construct and hence should not be simply abandoned in favor of more complex models. The results are largely consistent with our main findings. However, our additional analysis also reveals an insignificant, but substantial, decrease in reported course enjoyment, while the positive effects of effort are now statistically significant. This still aligns with our theoretical explanation, suggesting that students who perceive a subject as more demanding might consequently express less overall affinity for the course design. At a minimum, our findings align with \textcite{van_alten}, further suggesting an absence of evidence that students are inherently more satisfied with flipped learning environments.

Formal learning theories suggest that reducing procrastination and increasing students' self-concept in statistics should improve exam performance and knowledge retention. However, our findings indicate that the effects on these learning outcomes are neither positive nor statistically significant. The mechanism driving our null results is discussed further in the following section.

\section{Discussion of Mechanism}\label{s6_discussion}

For the flipped classroom cohort, individual-level usage data was collected using personal access credentials to videos and online quizzes. Video tracking data details viewing patterns such as if, when, and how often a specific 5-second segment of a video was watched. For online quizzes, participation, access time, submission time, and achieved points were recorded. Unique clickers enabled tracking of individual participation and correct responses during clicker sessions.

Using these data, we constructed four usage measures. \textit{Videos Due (VD)} quantifies timely video engagement as the product of the proportion of videos accessed before the exam and the average proportion of unique 5-second segments watched within the corresponding self-learning phases. \textit{Total Videos (TV)} measures overall video engagement as the product of the proportion of videos accessed and the average proportion of unique 5-second segments watched until the exam. \textit{Quiz Participation (QP)} assesses participation in online quizzes as the product of the share of quizzes attempted and the average proportion of questions answered per attempt. \textit{Active Classroom Sessions (ACS)} measures in-class participation as the product of the proportion of active sessions attended and the average proportion of relevant clicker questions answered during attendance. All usage measures range from zero to one, with a value of one indicating ideal engagement with this part of the curriculum and zero indicating no engagement.

The measures \textit{VD}, \textit{TV}, and \textit{QP} assess student behavior during knowledge acquisition, while \textit{ACS} monitors the knowledge deepening phase in active classroom sessions. However, note that \textit{TV} serves only as a complementary measure, as the flipped classroom design, introduced in Section \ref{s2_ins}, requires knowledge acquisition during self-learning phases for effective participation in subsequent knowledge-deepening activities. Thus, delaying self-learning phases, for example, postponing video watching, would substantially limit the benefits of active classroom sessions. In this sense, \textit{VD} can be viewed as the primary metric for assessing student engagement with this instructional format. \textit{QP} and \textit{ACS} can be considered equally important. \textit{QP} assesses lower-order cognitive skills and reveals fundamental misconceptions of knowledge acquired during self-learning, whereas \textit{ACS} fosters higher-order cognitive skills through active participation during classroom sessions.\footnote{Lower-order cognitive skills include recalling, understanding, and applying the learned material, whereas higher-order cognitive skills encompass advanced mental processes, including reasoning, critical thinking, problem-solving, and creating (see \textcites{lewis, anderson} for more details).}

\begin{figure}[!htbp]
    \caption{Usage Measures by Exam Performance}
    \label{fig:usage}
    \centering
    \includegraphics[width=0.9\linewidth]{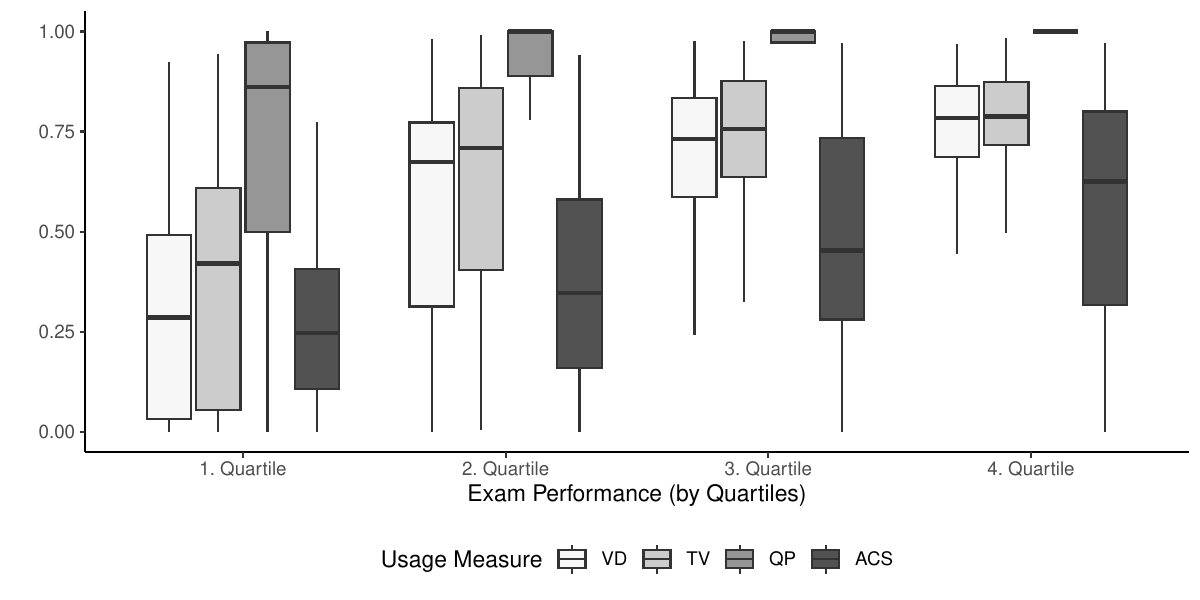}
    \floatfoot{\scriptsize\raggedright\emph{Note:} Students ($n = 196$) assigned based on exam points: 1. Quartile ($2 \leq \text{exam points} \leq 19.5$), 2. Quartile ($20 \leq \text{exam points} \leq 27$), 3. Quartile ($27.5 \leq \text{exam points} \leq 35.5$), 4. Quartile ($36 \leq \text{exam points} \leq 57$).}
\end{figure}

Figure \ref{fig:usage} presents the distribution of our four constructed usage measures by quartiles of exam points. In general, we observe a positive trend between usage and performance. For example, contrasting \textit{VD} between the first quartile and the fourth quartile suggests a strong association of exam performance with watching videos in time. In particular, average \textit{VD} increased from approximately 30\% in the first quartile to 75\% in the fourth. A comparison of \textit{VD} and \textit{TV} within quartiles indicates that the students in the first quartile exhibited the most catch-up behavior, yet their average \textit{TV}, roughly 40\%, remained significantly lower than the average for all other quartiles. Interestingly, for students in the fourth quartile, we observe an average \textit{TV} of only roughly 80\%. It is important to note that the high values we observe for \textit{QP} likely reflect the bonus point incentive to correctly answer questions in the quiz, with the upper three quartiles showing near-complete engagement, whereas the first quartile engages only in roughly 70\% of the quizzes. Finally, we observe substantially lower values for \textit{ACS} with larger dispersion for higher quartiles, potentially due to the requirement that students bring personal clickers to participate actively in classroom sessions, suggesting occasional nonparticipation in clicker polls.

The general low engagement in watching videos on time suggests that many students may not have adequately comprehended or retained the content necessary to fully benefit from self-testing and deepening of the knowledge in classroom sessions. This observed usage pattern offers a plausible explanation for the absence of statistically significant effects on learning outcomes. Furthermore, we find that students perceive themselves as more capable and procrastinate less, which could add to the problem that they did not engage properly with the course content. Since active and timely engagement is crucial for the success of the flipped classroom, our findings can be considered in line with the underlying learning-theoretical expectations.

Finally, as a side result, we note that the number of videos accessed, a frequently used proxy in flipped classroom literature, should be complemented with a rich set of covariates to remediate potential measurement error, when used as a substitute for more granular and time-specific video usage measures in assessing the effects of a flipped classroom (see Online Appendix \ref{app:usage_access} for details).

\section{Concluding Remarks}\label{s7_concl}

We used a cohort design and causal machine learning to analyze the effect of changing the instructional format of a large introductory statistics course for undergraduate students from lecture-based teaching to a flipped classroom.

In general, we did not find evidence that changing the instructional format affected exam performance, but we found statistically significant and meaningful effects on learning strategies and perceptions. Detailed usage data from the flipped cohort indicate that students did not use the instructional approach adequately, which explains the mechanism of our null results from a learning-theoretic point of view. Hence, our findings underscore the need to integrate complementary strategies to ensure timely and effective student engagement with the learning material when implementing flipped classrooms in higher education. 

Finally, we want to emphasize the clear advantage and usefulness of \textcite{chernozhukov}'s double/debiased machine learning (DML) compared to traditional estimation approaches. The combination of DML with appropriate machine learning techniques, such as the ridge regression employed herein to deal with the many highly correlated covariates and Likert items, offers a powerful and objective approach for future studies that face similar challenges.

\clearpage

% % Referenecs % % % % % % % % % % % %

\pagestyle{plain}

\renewcommand*{\bibfont}{\small}

\printbibliography 

\clearpage

% % % % % % % % % % % %Online Appendix% % % % % % % % % % % % % % % % % % % % %

\appendix

{\noindent\LARGE\textbf{Online Appendix}}

\section{Likert Scales in Questionnaire}
\label{app:likert_scales}

This section outlines the items collected for each scale in the questionnaires. All items were created based on  psychometrically validated scales, ensuring theoretical alignment and internal consistency. The following table presents the final scales, and cites relevant sources and original item numbers.

\begin{table}[!htbp]
    \footnotesize
	\caption{Likert Scales and Items used in Questionnaire}
    \label{tab:likert_scales}
	\begin{tabular}{llll}
        \toprule
        Scale & Citation & Items& Plus Modification(s) of...\\
        \midrule
        \underline{Perception of Statistics}&&\\
        \hspace{.1cm} Affection&\textcite{schau}&3, 15, 18, 28&3, 28\\
        \hspace{.1cm} Difficulty&\textcite{schau}&8, 22, 24, 30, 34, 36&\\
        \hspace{.1cm} Effort&\textcite{schau}&1, 2, 14, 27&27 ($2\times$)\\
        \hspace{.1cm} Extrinsic Motivation&\textcite{pintrich}&7, 11, 13, 30&\\
        \hspace{.1cm} Interest&\textcite{schau}&12, 20, 23, 29&\\
        \hspace{.1cm} Self-Concept&\textcite{schau}&5, 11, 31,32, 35&\\ 
        \hspace{.1cm} Value&\textcite{schau}&7, 10, 17, 25, 33&7, 9, 13\\
        \underline{Learning}&&&\\
        \hspace{.1cm} Critical Thinking&\textcite{pintrich}&38, 47, 51, 66, 71&\\
        \hspace{.1cm} Environment&\textcite{pintrich}&35, 43, 52, 70, 80&\\
        \hspace{.1cm} Procrastination&\textcite{patzelt}&1, 3, 6, 10, 12&\\
        \hspace{.1cm} Repetition&\textcite{pintrich}&39, 46, 59, 72& 62\\
        \hspace{.1cm} Peer&\textcite{pintrich}&34, 50& 45\\
        \hspace{.1cm} Self-Regulation&\textcite{pintrich}&33, 56, 57, 76, 78& 41 combined with 79\\
        \hspace{.1cm} Test-Anxiety&\textcite{pintrich}&3, 8, 14, 19, 28&\\
        \underline{Course Perception}&&&\\
        \hspace{.1cm} Boredom&\textcite{pekrun}&29, 42, 51, 61, 66&66\\
        \hspace{.1cm} Enjoyment&\textcite{pekrun}&1, 5, 11, 24, 41, 71, 76&\\
        \hspace{.1cm} Frustration&\textcite{pekrun}&14, 22, 55, 75&55\\
        \hspace{.1cm} Self-Concept&\textcite{pekrun}&4, 7, 37& 20\\
        \bottomrule
    \end{tabular}
\end{table}

\section{Item Analysis}
\label{app:item_analysis}

This section describes the item analysis performed to evaluate the psychometric properties of the scales and outlines the procedure for selecting the items.

Likert-item responses from both cohorts were pooled. The corresponding scales were analyzed separately. We simultaneously assessed multiple psychometric criteria. Internal consistency was assessed using Cronbach's $\alpha$ and McDonald's $\omega$ \parencites{cronbach, mcdonald}. To evaluate the assumption of tau-equivalence, we compared the model fit of a unidimensional confirmatory factor analysis (CFA) with equal loadings to that of a congeneric CFA model for each scale. The congeneric model allows items to load differently on a single latent factor, whereas the tau-equivalent model constrains all items to have the same association with the latent construct. Consistent with expectations for psychological scales, results revealed substantial deviations from tau-equivalence across nearly all scales, thus necessitating the use of McDonald's $\omega$ as the primary measure of reliability. Furthermore, item quality was assessed considering corrected item-total correlations and standardized congeneric CFA loadings. Results for the first and second questionnaire are presented in Table \ref{tab:item_analysis_sample_a} and \ref{tab:item_analysis_sample_c}.

\begin{ThreePartTable}
\footnotesize
\centering
\begin{TableNotes}
\scriptsize
\item \emph{Note:} Cronbach's $\alpha$ denotes the 95\% confidence interval constructed around the $\alpha$; CFA - RMSR (standardized) denotes the standardized root mean square residual after Confirmatory Factor Analysis (CFA); all measures reported were computed using the statistical software R \parencite{r_software} and the following packages: \textit{psych} \parencite{psych_software} (for Cronbach's $\alpha$, McDonald's $\omega$ and (corrected) Item-total correlation) and \textit{lavaan} \parencite{lavaan_software} (for CFA).
\end{TableNotes}
\begin{longtable}{llcccc}
\caption{Item Analysis for Likert Data from First Questionnaire}\label{tab:item_analysis_sample_a}\\
\toprule
Scale & Item & Cronbach's $\alpha$ & McDonald's $\omega$ & Item-total Corr. & CFA - RMSR \\
&&&&(corrected)&(standardized)\\
\midrule
\endfirsthead

\multicolumn{6}{c}{\tablename\ \thetable\ -- Continued from previous page} \\
\midrule
Scale & Item & Cronbach's $\alpha$ & McDonald's $\omega$ & Item-total Corr. & CFA - RMSR \\
&&&&(corrected)&(standardized)\\
\midrule
\endhead

\midrule
\multicolumn{6}{r}{Continued on next page} \\
\bottomrule
\endfoot

\midrule
\multicolumn{6}{r}{End of Table} \\
\bottomrule
\insertTableNotes
\endlastfoot

  Affection& 1 & (0.749, 0.812) & 0.784 & 0.416 & 0.087 \\ 
   &2 &  &  & 0.497 &  \\ 
   &3 &  &  & 0.552 &  \\ 
   &4 &  &  & 0.633 &  \\ 
   &5 &  &  & 0.631 &  \\ 
   &6 &  &  & 0.459 &  \\ 
  Difficulty &   1 & (0.638, 0.731) & 0.699 & 0.514 & 0.049 \\ 
   &   2 &  &  & 0.391 &  \\ 
   &   3 &  &  & 0.512 &  \\ 
   &   4 &  &  & 0.542 &  \\ 
   &   5 &  &  & 0.281 &  \\ 
   &   6 &  &  & 0.271 &  \\ 
  Effort &   1 & (0.843, 0.883) & 0.867 & 0.689 & 0.071 \\ 
   &   2 &  &  & 0.731 &  \\ 
   &   3 &  &  & 0.720 &  \\ 
   &   4 &  &  & 0.596 &  \\ 
   &   5 &  &  & 0.600 &  \\ 
   &   6 &  &  & 0.629 &  \\ 
  Extrinsic Motivation &   1 & (0.737, 0.807) & 0.806 & 0.617 & 0.037 \\ 
   &   2 &  &  & 0.675 &  \\ 
   &   3 &  &  & 0.614 &  \\ 
   &   4 &  &  & 0.498 &  \\ 
  Interest &   1 & (0.840, 0.884) & 0.876 & 0.613 & 0.029 \\ 
   &   2 &  &  & 0.817 &  \\ 
   &   3 &  &  & 0.762 &  \\ 
   &   4 &  &  & 0.694 &  \\ 
  Self-Concept &   1 & (0.750, 0.816) & 0.797 & 0.565 & 0.067 \\ 
   &   2 &  &  & 0.653 &  \\ 
   &   3 &  &  & 0.509 &  \\ 
   &   4 &  &  & 0.665 &  \\ 
   &   5 &  &  & 0.445 &  \\ 
  Value &   1 & (0.792, 0.843) & 0.832 & 0.562 & 0.054 \\ 
   &   2 &  &  & 0.586 &  \\ 
   &   3 &  &  & 0.723 &  \\ 
   &   4 &  &  & 0.660 &  \\ 
   &   5 &  &  & 0.289 &  \\ 
   &   6 &  &  & 0.658 &  \\ 
   &   7 &  &  & 0.299 &  \\ 
   &   8 &  &  & 0.580 &  \\ 
  Critical Thinking &   1 & (0.677, 0.762) & 0.724 & 0.434 & 0.056 \\ 
   &   2 &  &  & 0.444 &  \\ 
   &   3 &  &  & 0.548 &  \\ 
   &   4 &  &  & 0.472 &  \\ 
   &   5 &  &  & 0.503 &  \\ 
  Environment &   1 & (0.716, 0.790) & 0.760 & 0.515 & 0.029 \\ 
   &   2 &  &  & 0.587 &  \\ 
   &   3 &  &  & 0.537 &  \\ 
   &   4 &  &  & 0.528 &  \\ 
   &   5 &  &  & 0.454 &  \\ 
  Procrastination &   1 & (0.803, 0.854) & 0.835 & 0.655 & 0.030 \\ 
   &   2 &  &  & 0.763 &  \\ 
   &   3 &  &  & 0.765 &  \\ 
   &   4 &  &  & 0.437 &  \\ 
   &   5 &  &  & 0.520 &  \\ 
  Repetition &   1 & (0.476, 0.616) & 0.558 & 0.354 & 0.067 \\ 
   &   2 &  &  & 0.381 &  \\ 
   &   3 &  &  & 0.280 &  \\ 
   &   4 &  &  & 0.334 &  \\ 
  Peer &   1 & (0.748, 0.821) & 0.812 & 0.453 & 0.000 \\ 
   &   2 &  &  & 0.743 &  \\ 
   &   3 &  &  & 0.702 &  \\ 
  Self-Regulation &   1 & (0.677, 0.760) & 0.767 & 0.410 & 0.066 \\ 
   &   2 &  &  & 0.586 &  \\ 
   &   3 &  &  & 0.683 &  \\ 
   &   4 &  &  & 0.596 &  \\ 
   &   5 &  &  & 0.258 &  \\ 
   &   6 &  &  & 0.226 &  \\ 
   &   7 &  &  & 0.449 &  \\ 
  Test-Anxiety &   1 & (0.803, 0.855) & 0.836 & 0.599 & 0.038 \\ 
   &   2 &  &  & 0.606 &  \\ 
   &   3 &  &  & 0.675 &  \\ 
   &   4 &  &  & 0.772 &  \\ 
   &   5 &  &  & 0.502 &  \\ 
\end{longtable}
\end{ThreePartTable}

\begin{ThreePartTable}
\footnotesize
\centering
\begin{TableNotes}
\scriptsize
\item \emph{Note:} Cronbach's $\alpha$ denotes the 95\% confidence interval constructed around the $\alpha$; CFA - RMSR (standardized) denotes the standardized root mean square residual after Confirmatory Factor Analysis (CFA); all measures reported were computed using the statistical software R \parencite{r_software} and the following packages: \textit{psych} \parencite{psych_software} (for Cronbach's $\alpha$, McDonald's $\omega$ and (corrected) Item-total correlation) and \textit{lavaan} \parencite{lavaan_software} (for CFA).
\end{TableNotes}
\begin{longtable}{llcccc}
\caption{Item Analysis for Likert Data from Second Questionnaire}\label{tab:item_analysis_sample_c}\\
\toprule
Scale & Item & Cronbach's $\alpha$ & McDonald's $\omega$ & Item-total Corr. & CFA - RMSR \\
&&&&(corrected)&(standardized)\\
\midrule
\endfirsthead

\multicolumn{6}{c}{\tablename\ \thetable\ -- Continued from previous page} \\
\midrule
Scale & Item & Cronbach's $\alpha$ & McDonald's $\omega$ & Item-total Corr. & CFA - RMSR \\
&&&&(corrected)&(standardized)\\
\midrule
\endhead

\midrule
\multicolumn{6}{r}{Continued on next page} \\
\bottomrule
\endfoot

\midrule
\multicolumn{6}{r}{End of Table} \\
\bottomrule
\insertTableNotes
\endlastfoot

  Affection &   1 & (0.677, 0.784) & 0.756 & 0.426 & 0.092 \\ 
   &   2 &  &  & 0.561 &  \\ 
   &   3 &  &  & 0.522 &  \\ 
   &   4 &  &  & 0.633 &  \\ 
   &   5 &  &  & 0.534 &  \\ 
   &   6 &  &  & 0.176 &  \\ 
  Difficulty &   1 & (0.620, 0.746) & 0.690 & 0.513 & 0.057 \\ 
   &   2 &  &  & 0.207 &  \\ 
   &   3 &  &  & 0.457 &  \\ 
   &   4 &  &  & 0.550 &  \\ 
   &   5 &  &  & 0.420 &  \\ 
   &   6 &  &  & 0.329 &  \\ 
  Effort &   1 & (0.561, 0.722) & 0.710 & 0.445 & 0.000 \\ 
   &   2 &  &  & 0.615 &  \\ 
   &   3 &  &  & 0.340 &  \\ 
  Extrinsic Motivation &   1 & (0.725, 0.822) & 0.798 & 0.668 & 0.052 \\ 
   &   2 &  &  & 0.635 &  \\ 
   &   3 &  &  & 0.566 &  \\ 
   &   4 &  &  & 0.490 &  \\ 
  Interest &   1 & (0.828, 0.890) & 0.866 & 0.585 & 0.034 \\ 
   &   2 &  &  & 0.778 &  \\ 
   &   3 &  &  & 0.743 &  \\ 
   &   4 &  &  & 0.726 &  \\ 
  Self-Concept &   1 & (0.787, 0.860) & 0.831 & 0.657 & 0.031 \\ 
   &   2 &  &  & 0.704 &  \\ 
   &   3 &  &  & 0.624 &  \\ 
   &   4 &  &  & 0.666 &  \\ 
   &   5 &  &  & 0.465 &  \\ 
  Value &   1 & (0.789, 0.858) & 0.837 & 0.455 & 0.073 \\ 
   &   2 &  &  & 0.687 &  \\ 
   &   3 &  &  & 0.712 &  \\ 
   &   4 &  &  & 0.670 &  \\ 
   &   5 &  &  & 0.412 &  \\ 
   &   6 &  &  & 0.588 &  \\ 
   &   7 &  &  & 0.284 &  \\ 
   &   8 &  &  & 0.614 &  \\ 
  Critical Thinking &   1 & (0.752, 0.837) & 0.796 & 0.477 & 0.040 \\ 
   &   2 &  &  & 0.587 &  \\ 
   &   3 &  &  & 0.586 &  \\ 
   &   4 &  &  & 0.601 &  \\ 
   &   5 &  &  & 0.623 &  \\ 
  Environment  &   1 & (0.522, 0.686) & 0.636 & 0.394 & 0.068 \\ 
   &   2 &  &  & 0.524 &  \\ 
   &   3 &  &  & 0.365 &  \\ 
   &   4 &  &  & 0.233 &  \\ 
   &   5 &  &  & 0.319 &  \\ 
  Procrastination &   1 & (0.738, 0.827) & 0.796 & 0.684 & 0.022 \\ 
   &   2 &  &  & 0.678 &  \\ 
   &   3 &  &  & 0.666 &  \\ 
   &   4 &  &  & 0.372 &  \\ 
   &   5 &  &  & 0.415 &  \\ 
  Repetition  &   1 & (0.652, 0.770) & 0.728 & 0.502 & 0.078 \\ 
   &   2 &  &  & 0.585 &  \\ 
   &   3 &  &  & 0.499 &  \\ 
   &   4 &  &  & 0.409 &  \\ 
   &   5 &  &  & 0.387 &  \\ 
  Peer &   1 & (0.724, 0.825) & 0.800 & 0.447 & 0.000 \\ 
   &   2 &  &  & 0.747 &  \\ 
   &   3 &  &  & 0.661 &  \\ 
  Self-Regulation  &   1 & (0.592, 0.728) & 0.704 & 0.330 & 0.072 \\ 
   &   2 &  &  & 0.530 &  \\ 
   &   3 &  &  & 0.602 &  \\ 
   &   4 &  &  & 0.590 &  \\ 
   &   5 &  &  & 0.160 &  \\ 
   &   6 &  &  & 0.215 &  \\ 
   &   7 &  &  & 0.282 &  \\ 
  Test-Anxiety &   1 & (0.836, 0.891) & 0.866 & 0.618 & 0.027 \\ 
   &   2 &  &  & 0.656 &  \\ 
   &   3 &  &  & 0.753 &  \\ 
   &   4 &  &  & 0.762 &  \\ 
   &   5 &  &  & 0.640 &  \\ 
  Boredom  &   1 & (0.921, 0.948) & 0.936 & 0.715 & 0.029 \\ 
   &   2 &  &  & 0.821 &  \\ 
   &   3 &  &  & 0.815 &  \\ 
   &   4 &  &  & 0.764 &  \\ 
   &   5 &  &  & 0.864 &  \\ 
   &   6 &  &  & 0.863 &  \\ 
  Enjoyment &   1 & (0.880, 0.920) & 0.904 & 0.760 & 0.058 \\ 
   &   2 &  &  & 0.473 &  \\ 
   &   3 &  &  & 0.656 &  \\ 
   &   4 &  &  & 0.686 &  \\ 
   &   5 &  &  & 0.817 &  \\ 
   &   6 &  &  & 0.810 &  \\ 
   &   7 &  &  & 0.756 &  \\ 
  Frustration &   1 & (0.866, 0.912) & 0.892 & 0.736 & 0.062 \\ 
   &   2 &  &  & 0.709 &  \\ 
   &   3 &  &  & 0.731 &  \\ 
   &   4 &  &  & 0.761 &  \\ 
   &   5 &  &  & 0.736 &  \\ 
  Course Self-Concept &   1 & (0.926, 0.951) & 0.939 & 0.782 & 0.021 \\ 
   &   2 &  &  & 0.853 &  \\ 
   &   3 &  &  & 0.814 &  \\ 
   &   4 &  &  & 0.766 &  \\ 
   &   5 &  &  & 0.850 &  \\ 
   &   6 &  &  & 0.832 &  \\ 

\end{longtable}
\end{ThreePartTable}

The items were selected and refined on the basis of psychometric properties and theoretical relevance. Items with corrected item-total correlations below 0.3 were flagged \parencites{nunnally1, nunnally2, boateng}, indicating a weak association with the overall scale \parencite{devellis}. Despite an acceptable global model fit indicated by the standardized root mean square residual (RMSR) \parencites{hu1999}, items with standardized factor loadings below 0.4 in the congeneric Confirmatory Factor Analysis (CFA) model were considered for exclusion to improve the validity of the construct \parencite{stevens}. For flagged items, the impact of removal on omega reliability was assessed and item wording was reviewed for clarity. Items were dropped if their removal increased omega reliability or if their wording was vague or unclear \parencites{clark1995, dunn_alpha, devellis}.

This procedure led to the removal of the following:
\begin{table}[htbp!]
    \centering
    \begin{tabular}{lll}
    \toprule
    Scale&Item(s) removed& Note\\
    \midrule
    Difficulty&5, 6& \\
    Value&5, 7& \\
    Self-Regulation&5, 6& \\
    Repetition&5&only in second questionnaire,\\
    &&did not theoretically fit the scope of the scale\\
    \bottomrule
    \end{tabular}
\end{table}

Moreover, since reliability estimates that exceed the value of 0.7 are generally considered acceptable \parencites{nunnally2, mcneish}, scales for which McDonald's $\omega$ after item selection remained larger than this threshold were considered to demonstrate enough internal consistency to represent a common latent variable. Thus, for all scales presented, except repetition, the respective Likert items were aggregated by computing scale means, or further assessed to process with Principal Component Analysis (PCA). In Online Appendix \ref{app:pca}, we assess the adequacy of our data for dimension reduction using PCA.

Although Likert means are widely used and intuitively understandable for dimension reduction in applied research, their ability to accurately measure an underlying latent construct is compromised when tau-equivalence is violated. Recognizing these limitations, we extended our analysis by using Principal Component Analysis (PCA). Although not directly derived from psychometric theory, PCA offers a data-driven method that weights items based on shared variance. This approach addresses some of the inherent disadvantages of Likert means, potentially capturing more of the underlying latent construct. However, depending on the scale, PCA may suggest extracting multiple principal components. When these components serve as covariates in the estimation procedure, this does not present an issue. Conversely, if they are used as dependent variables, this multi-component extraction poses a problem for representing a single latent construct. To provide a psychometrically robust alternative in such scenarios, we also present results based on Item Response Theory (IRT) in Online Appendix \ref{app:irt_results}.

For the subsample in the second questionnaire, item analysis revealed altered reliability, particularly for the environment scale, which showed reduced internal consistency with a McDonald's $\omega$ of 0.636 (see Table \ref{tab:item_analysis_sample_c}).

\section{Principal Component Analysis} 
\label{app:pca}

This section outlines the procedures used to assess the adequacy of the data for dimension reduction and the number of principal components to retain as covariates in the estimation approach. 
Polychoric correlation matrices were computed for each scale using the Likert item responses from both cohorts. Based on these matrices, adequacy was assessed with the Kaiser-Meyer-Olkin measure of sampling adequacy and Bartlett's test of sphericity \parencites{bartlett, kaiser1970, dziuban, kaiser_rice}.

Moreover, the determinant of each polychoric correlation matrix was examined as an indicator of multicollinearity and to assess matrix invertibility \parencite{shrestha}. In addition to the Jolliffe Criterion \parencite{jolliffe1972}, several factor retention criteria were applied to determine the optimal number of principal components based on PCA eigenvalues. These include scree plots, the Kaiser-Guttman Criterion, the Empirical Kaiser Criterion, and results from parallel analysis. For more details on these retention criteria, see, among others, \textcites{guttman, horn, cattell, zwick,  crawford, braeken, auerswald}. Although developed for factor analysis, these criteria are commonly applied in principal component analysis, given their reliance on eigenvalue decomposition \parencites[Chapter 6]{zwick,jolliffe}.

Table \ref{tab:diagnostics} gives an overview of the test results and the number of components to retain based on each criteria. For all scales, the KMO is acceptable, the Bartlett is significant, and the determinant score showed values larger than the critical threshold, indicating the adequacy for dimension reduction and matrix invertibility. Moreover, for all scales, most retention criteria indicate that a single component should be retained for subsequent analysis.

\begin{ThreePartTable}
\footnotesize
\centering
\begin{TableNotes}
\scriptsize
\item \emph{Note:} Kaiser-Meyer-Olkin (KMO) ; Bartlett's Test for Sphericity (Bartlett) ; Determinant of the correlation matrix ; Empirical Kaiser Criterion (EKC) ; Jolliffe Criterium (JC) ; Kaiser-Guttman Criterion (KGC) ; Parallel Analysis (PA) ; all measures reported were computed using the statistical software R \parencite{r_software} and the following packages 
\textit{EFA.dimensions} \parencite{efadim_software} (for polychoric correlation matrices), \textit{EFAtools} \parencite{efatools_software} (for assessing dimension reduction suitability and factor retention criteria).

\end{TableNotes}

\begin{longtable}{m{4cm} >{\centering\arraybackslash}m{8cm}} 
\caption{Diagnostics--Dimension Reduction}\label{tab:diagnostics}\\
\toprule
Scale & First Questionnaire \\
\midrule
\endfirsthead

\multicolumn{2}{c}{\tablename\ \thetable\ -- Continued from previous page} \\
\midrule
Scale & First Questionnaire \\ 
\midrule
\endhead

\midrule
\multicolumn{2}{r}{Continued on next page} \\
\bottomrule
\endfoot

\midrule
\multicolumn{2}{r}{End of Table} \\
\bottomrule
\insertTableNotes
\endlastfoot

Affection & \includegraphics[width = 8cm, height = 8cm]{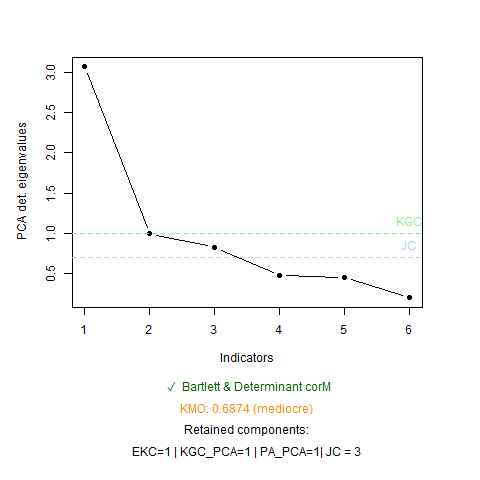} \\ 
Difficulty &  \includegraphics[width = 8cm, height = 8cm]{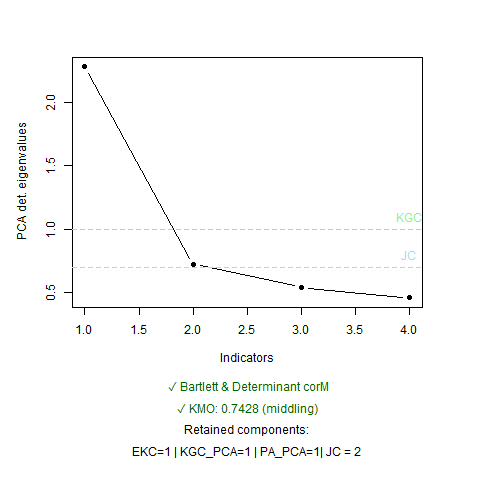} \\
Effort & \includegraphics[width = 8cm, height = 8cm]{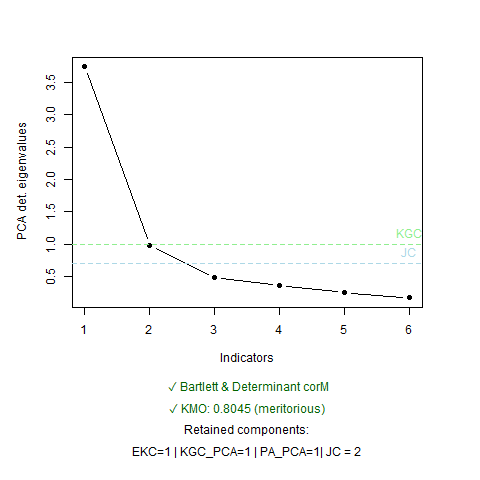} \\
\makecell[ll]{Extrinsic\\Motivation} & \includegraphics[width = 8cm, height = 8cm]{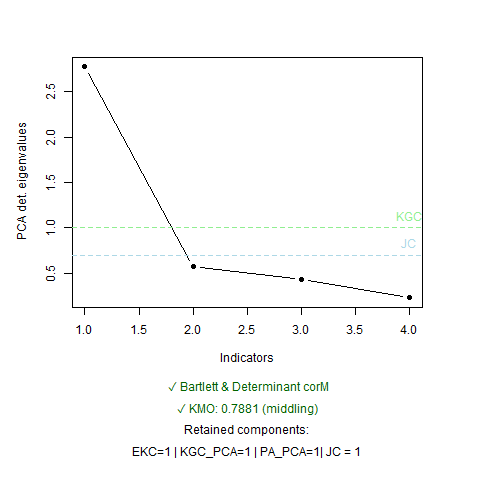} \\
Interest & \includegraphics[width = 8cm, height = 8cm]{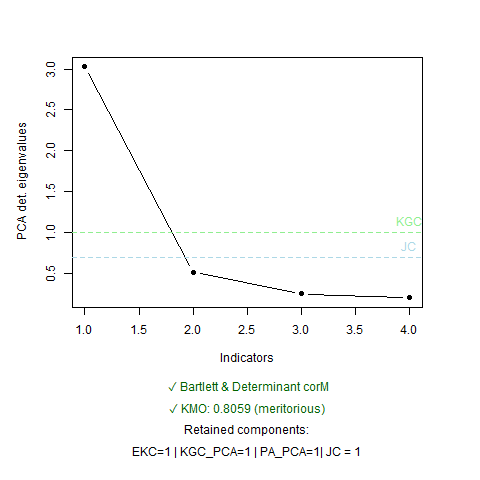} \\  
Self-Concept & \includegraphics[width = 8cm, height = 8cm]{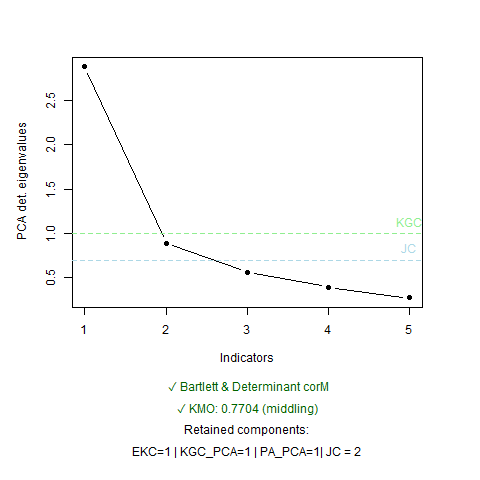} \\
Value &  \includegraphics[width = 8cm, height = 8cm]{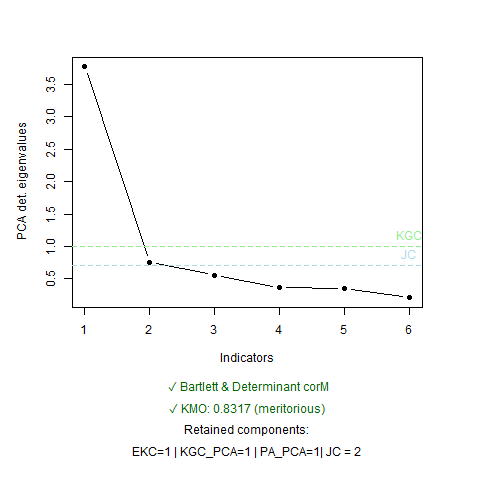} \\
\makecell[ll]{Critical\\Thinking} &  \includegraphics[width = 8cm, height = 8cm]{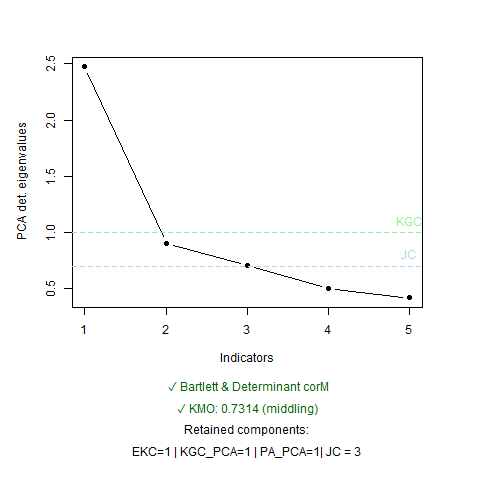} \\
Environment & \includegraphics[width = 8cm, height = 8cm]{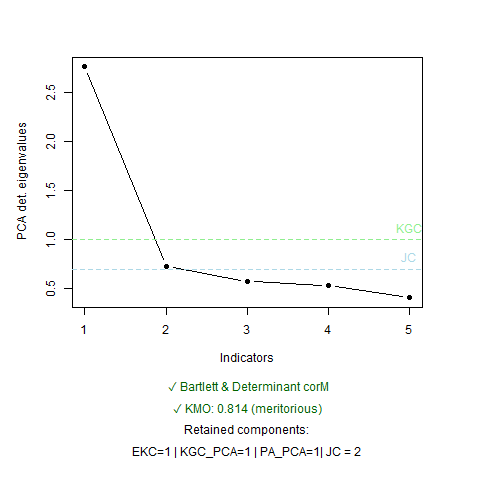} \\
Procrastination & \includegraphics[width = 8cm, height = 8cm]{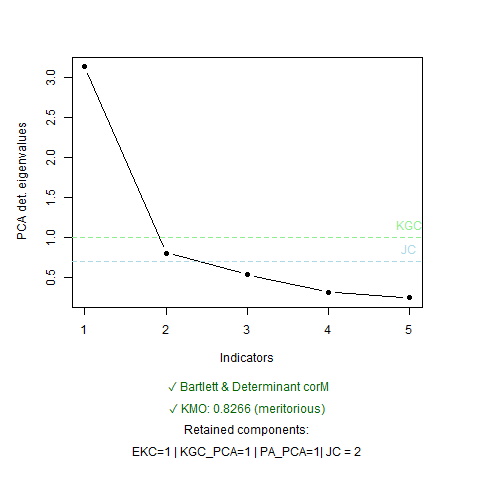} \\
Peer &  \includegraphics[width = 8cm, height = 8cm]{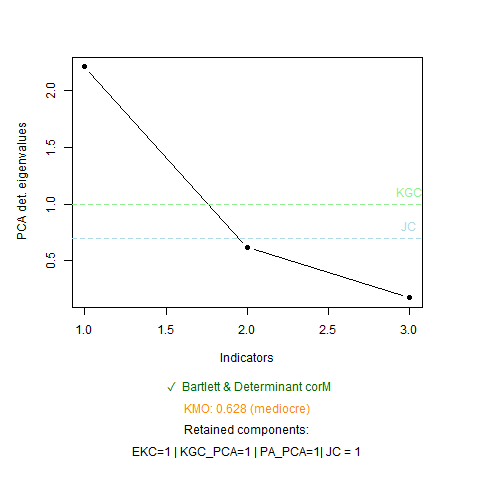} \\
Self-Regulation & \includegraphics[width = 8cm, height = 8cm]{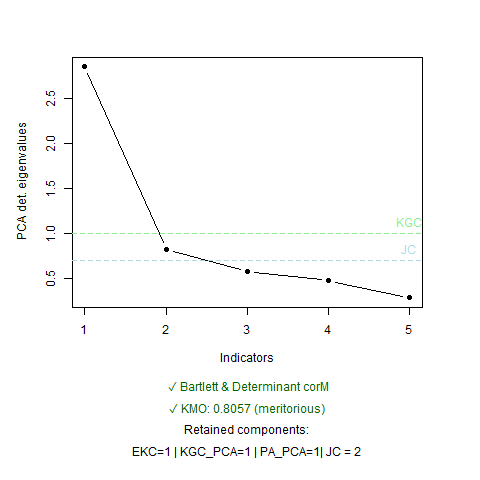} \\
Test-Anxiety &   \includegraphics[width = 8cm, height = 8cm]{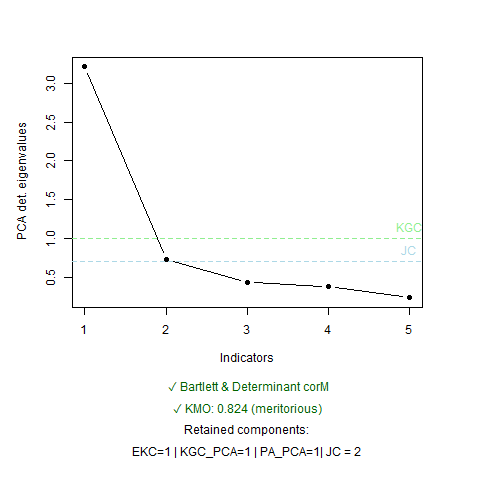} \\ 
   
\end{longtable}
\end{ThreePartTable}

\clearpage

\section{IRT Results} \label{app:irt_results}

\begin{table}[!htbp]
	\centering
	\footnotesize
    \begin{threeparttable}
	\caption{Estimated Average Treatment Effects (ATEs) Replacing Means by IRT Trait Estimates}
    \label{tab:estimated_ates_irt}
	\begin{tabular}{llccccc}
		\toprule
		&Outcome variable & \multicolumn{5}{c}{Estimated ATE} \\
        \cmidrule(lr){3-7}
        && OLS (1) & OLS (2) & OLS (3) & DML (1) & DML (2)\\
		\midrule
		\textit{Sample C}&\underline{Perception of Statistics} &  &  & &&  \\[0.5em]
		($n_{C1} = 96$ / $n_{C2} = 130$)&\hspace{.1cm} Affection        &  0.387$^{**}$ &  0.212 &  0.211 &  0.262$^{**}$ &  0.255$^{**}$ \\ 
        && (0.167) & (0.134) & (0.135) & (0.113) & (0.120) \\ 
        &\hspace{.1cm} Difficulty        &  0.005 &  0.007 & -0.014 & -0.050 & -0.055 \\ 
        && (0.150) & (0.129) & (0.120) & (0.109) & (0.115) \\ 
        &\hspace{.1cm} Effort            &  0.416$^{**}$ &  0.388$^{**}$ &  0.392$^{***}$ &  0.310$^{**}$ &  0.312$^{**}$ \\ 
        && (0.173) & (0.152) & (0.149) & (0.126) & (0.131) \\ 
        &\hspace{.1cm} Extrinsic Motivation  & -0.165 & -0.127 & -0.188 & -0.070 & -0.077 \\ 
        && (0.168) & (0.142) & (0.147) & (0.115) & (0.122) \\ 
        &\hspace{.1cm} Interest          & -0.093 & -0.105 & -0.166 & -0.109 & -0.120 \\ 
        && (0.172) & (0.147) & (0.134) & (0.127) & (0.132) \\ 
		&\hspace{.1cm} Self-Concept      &  0.308 &  0.308$^{**}$ &  0.265$^{*}$ &  0.308$^{**}$ &  0.304$^{**}$ \\ 
        && (0.189) & (0.142) & (0.149) & (0.120) & (0.126) \\ 
		&\hspace{.1cm} Value             & -0.130 & -0.129 & -0.148 & -0.018 &  0.002 \\ 
        && (0.180) & (0.149) & (0.141) & (0.124) & (0.130) \\ 
		&\underline{Learning}              & & & & & \\[0.5em]
        &\hspace{.1cm} Critical Thinking &  0.016 &  0.005 & -0.024 & -0.067 & -0.062 \\ 
        && (0.158) & (0.128) & (0.127) & (0.119) & (0.126) \\ 
		&\hspace{.1cm} Environment       & -0.048 & -0.056 & -0.103 & -0.017 & -0.022 \\ 
        && (0.143) & (0.133) & (0.122) & (0.109) & (0.115) \\ 
		&\hspace{.1cm} Procrastination   & -0.129 & -0.235$^{*}$ & -0.184 & -0.217$^{*}$ & -0.229$^{*}$ \\ 
        && (0.152) & (0.127) & (0.124) & (0.118) & (0.123) \\
        &\hspace{.1cm} Peer        & -0.198 & -0.290$^{**}$ & -0.253$^{*}$ & -0.268$^{**}$ & -0.272$^{**}$ \\ 
        && (0.192) & (0.134) & (0.142) & (0.124) & (0.130) \\ 
        &\hspace{.1cm} Self-Regulation   &  0.315$^{**}$ &  0.267$^{**}$ &  0.238$^{*}$ &  0.189 &  0.175 \\ 
        && (0.155) & (0.129) & (0.127) & (0.116) & (0.121) \\ 
		&\hspace{.1cm} Test-Anxiety     & -0.082 & -0.118 & -0.098 & -0.149 & -0.148 \\ 
        && (0.156) & (0.111) & (0.110) & (0.107) & (0.119) \\ 
		&\underline{Course Perception}   & & & & & \\[0.5em]
        &\hspace{.1cm} Boredom     &  0.155 & -0.042 & -0.066 & -0.060 & -0.059 \\ 
        && (0.176) & (0.155) & (0.154) & (0.128) & (0.133) \\ 
		&\hspace{.1cm} Enjoyment  & -0.389$^{**}$ & -0.207 & -0.204 & -0.186 & -0.190 \\ 
        && (0.168) & (0.153) & (0.154) & (0.126) & (0.131) \\ 
        &\hspace{.1cm} Frustration & -0.048 & -0.113 & -0.129 & -0.096 & -0.100 \\ 
        && (0.183) & (0.145) & (0.148) & (0.124) & (0.129) \\ 
		&\hspace{.1cm} Self-Concept &  0.192 &  0.332$^{**}$ &  0.321$^{**}$ &  0.323$^{**}$ &  0.326$^{**}$ \\ 
        && (0.182) & (0.153) & (0.159) & (0.126) & (0.132) \\ 
		\bottomrule
	\end{tabular}
    \begin{tablenotes}
        \item \emph{Note:} Robust standard errors in parentheses; DML uses ridge regression/classification to estimate $g(d_{i}, x_{i})$ and $m(x_{i})$ in \eqref{eq:model}; tuning parameters for ridge regression/classification are determined by ten-fold cross-validation; DML results are based on five-fold cross-fitting and 100 repetitions.
        \item $^{*}$p$<$0.1; $^{**}$p$<$0.05; $^{***}$p$<$0.01 
    \end{tablenotes}
    \end{threeparttable}
\end{table}

\clearpage

\section{Video Usage vs. Access} \label{app:usage_access}

\begin{figure}[!htbp]
 \centering
 \begin{minipage}{0.48\textwidth}
 \centering
 \includegraphics[width=\linewidth]{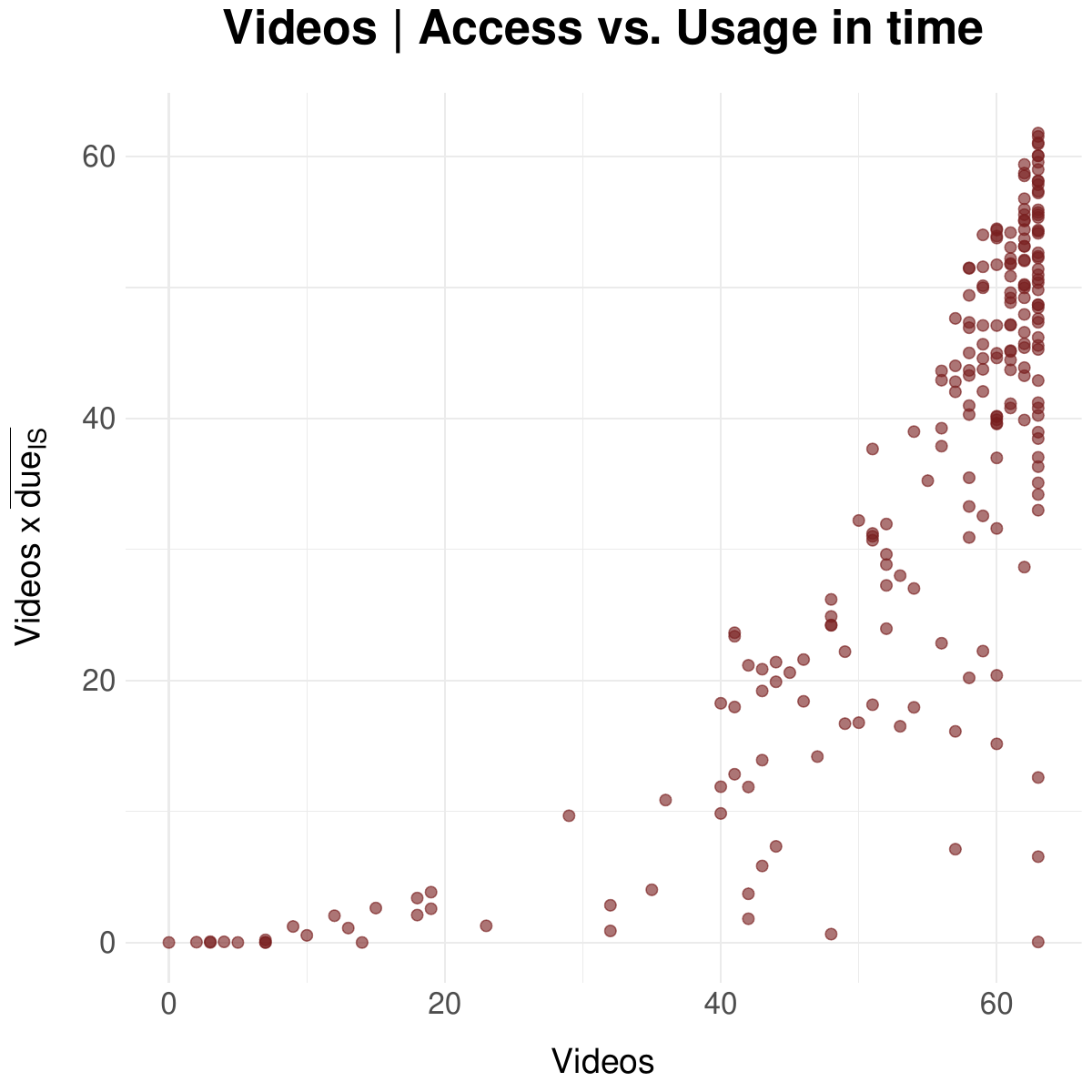}
 \end{minipage}%
 \hfill
 \begin{minipage}{0.48\textwidth}
 \centering
 \includegraphics[width=\linewidth]{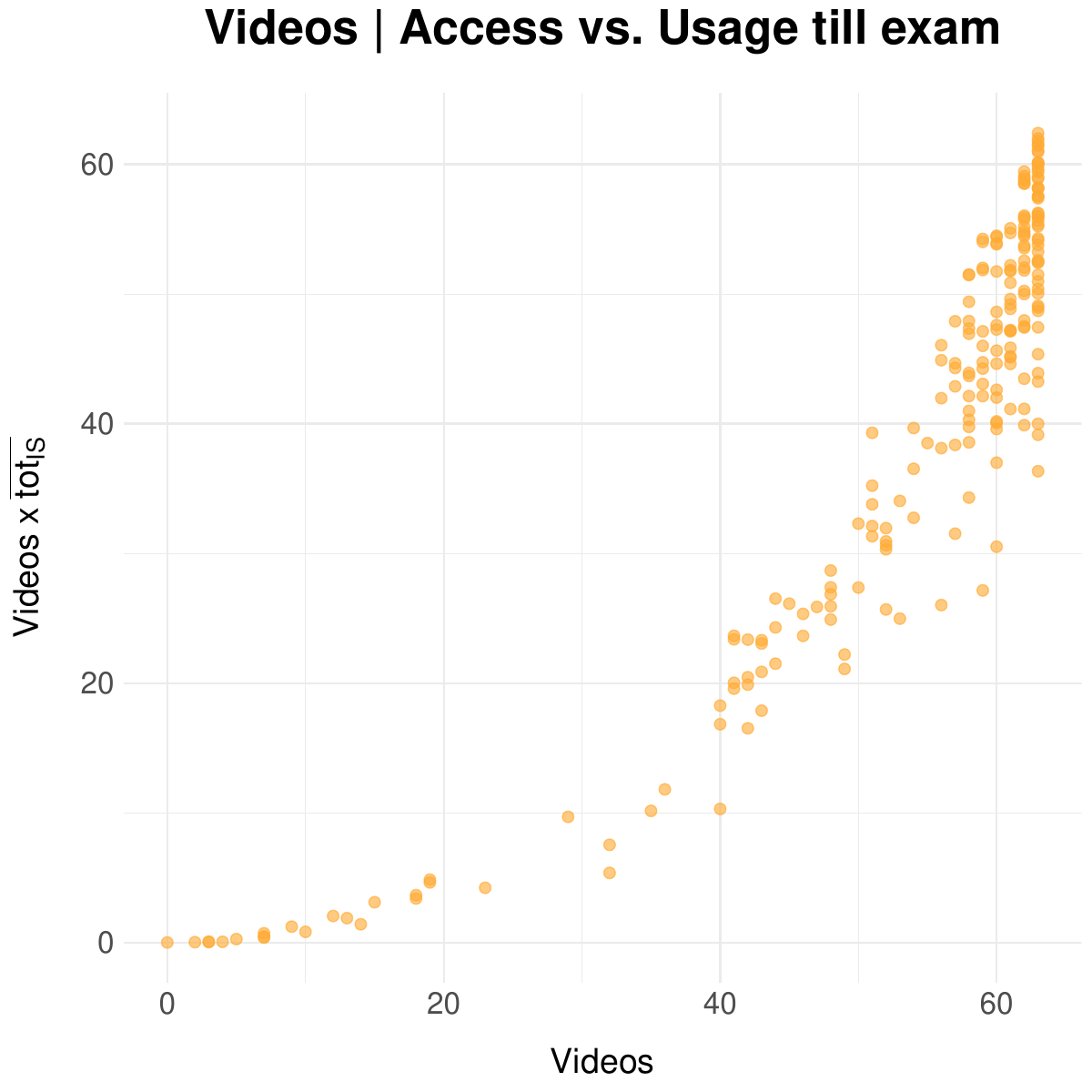}
 \captionof{figure}{Video Usage vs. Accesses}
  \label{fig:usageaccessvideo}
 \end{minipage}
\end{figure}

The video usage measures developed in this study can, moreover, serve as benchmarks for evaluating the quality of a frequently used proxy for timely video engagement in the flipped classroom literature, that is, the number of videos accessed by a student. To assess this, Figure \ref{fig:usageaccessvideo} provides scatterplots that map the total number of videos accessed by each student against this value times the average share of unique 5-second video segments each student watched in time before the respective classroom session took place $\big(\overline{\text{due}_{IS}}\big)$ or in general until the exam $\big(\overline{\text{tot}_{IS}}\big)$.

The left scatterplot provides information on timely video watching. It indicates that if one only controls for video access in the estimation procedure to reflect students’ video engagement during self-learning phases, one would consider students who accessed the same number of videos equally prepared and those who accessed more videos better prepared; which is not necessarily the case. The right scatterplot, on the other hand, reflects video usage until the exam. It indicates that the correlation of access and usage until the exam increases, suggesting that some students tried to catch up with missed content. Despite the increased association, the design of flipped classrooms implies that access alone remains insufficient to measure video engagement. This is the case because it does not reflect timely and conscientious engagement in this instructional framework. Students in a flipped classroom are only expected to gain additional lower-level cognitive skills through video watching after the due date of the respective classroom session but are not able to test and deepen this content to develop higher-level cognitive skills in the intended way. 

We thus recommend that if only video accesses can be used in assessing video usage in a flipped classroom, the estimation procedure should be extended by a rich set of covariates that include proxies for students' general conscientiousness and timeliness, and thus ensure that the core component in the flipped curriculum, namely, timely and sufficient video engagement is assessed adequately and not distorted by measurement error.

\end{document}